\ifx\compilefullpaper\undefined
\documentclass[10pt, twocolumn, aps, nofootinbib, longbibliography, nobibnotes, groupedaddress]{revtex4-1} 
\usepackage{amsfonts, amssymb, amsmath, amsthm}
\usepackage{siunitx} 
\usepackage{latexsym}
\usepackage[tracking=smallcaps]{microtype}	
\usepackage{url}
\usepackage{color}
\definecolor{DarkGray}{rgb}{0.1,0.1,0.5}
\definecolor{curgray}{rgb}{0.5,0.5,0.5}
\definecolor{curblue}{rgb}{0.04,0.11,0.64}
\definecolor{curpurple}{rgb}{0.65,0.16,0.58}
\definecolor{curorange}{rgb}{1,0.32,0}

\PassOptionsToPackage{hyphens}{url}

\usepackage[colorlinks=true,breaklinks=true, linkcolor=DarkGray,citecolor=DarkGray,urlcolor=DarkGray]{hyperref}	

\usepackage{graphicx}
\usepackage{subfigure}

\newcommand{\ket}[1]{{|#1\rangle}}

\newcommand{\binomial}[2]{\ensuremath{\left(\begin{smallmatrix}#1 \\ #2 \end{smallmatrix}\right)}}

\def\cC {{\mathcal C}}
\def\D {{\mathcal D}}

\newcounter{sprows}

\newlength{\spheight}
\newlength{\spraise}

\newlength{\commentslength}

\newcommand{\rem}[1]{}

\newtheorem{theorem}{Theorem}
\newtheorem{lemma}[theorem]{Lemma}
\newtheorem{corollary}[theorem]{Corollary}

\newtheorem{definition}[theorem]{Definition}

\newfont{\subsubsecfnt}{ptmri8t at 11pt}
\renewcommand{\subparagraph}[1]{\smallskip{\subsubsecfnt #1.}}

\newcommand{\eqnref}[1]{\hyperref[#1]{{(\ref*{#1})}}}
\newcommand{\thmref}[1]{\hyperref[#1]{{Theorem~\ref*{#1}}}}
\newcommand{\lemref}[1]{\hyperref[#1]{{Lemma~\ref*{#1}}}}
\newcommand{\corref}[1]{\hyperref[#1]{{Corollary~\ref*{#1}}}}
\newcommand{\defref}[1]{\hyperref[#1]{{Definition~\ref*{#1}}}}
\newcommand{\secref}[1]{\hyperref[#1]{{Sec.~\ref*{#1}}}}
\newcommand{\figref}[1]{\hyperref[#1]{{Fig.~\ref*{#1}}}}  
\newcommand{\figureref}[1]{\hyperref[#1]{{Figure~\ref*{#1}}}}  
\newcommand{\tabref}[1]{\hyperref[#1]{{Table~\ref*{#1}}}}
\newcommand{\remref}[1]{\hyperref[#1]{{Remark~\ref*{#1}}}}
\newcommand{\appref}[1]{\hyperref[#1]{{Appendix~\ref*{#1}}}}
\newcommand{\claimref}[1]{\hyperref[#1]{{Claim~\ref*{#1}}}}
\newcommand{\factref}[1]{\hyperref[#1]{{Fact~\ref*{#1}}}}
\newcommand{\propref}[1]{\hyperref[#1]{{Proposition~\ref*{#1}}}}
\newcommand{\exampleref}[1]{\hyperref[#1]{{Example~\ref*{#1}}}}
\newcommand{\conjref}[1]{\hyperref[#1]{{Conjecture~\ref*{#1}}}}

\allowdisplaybreaks[1]

\def\COLOR{}
\ifdefined\COLOR

\else

\fi

\definecolor{Cayenne}{rgb}{0.5,0,0}
\definecolor{Midnight}{rgb}{0,0,0.5}
\definecolor{Plum}{rgb}{0.5,0,0.5}
\definecolor{Teal}{rgb}{0,0.5,0.5}
\definecolor{Clover}{rgb}{0,0.5,0}
\definecolor{Maroon}{rgb}{0.5,0,0.25}
\definecolor{Ocean}{rgb}{0,0.25,0.5}
\definecolor{Tangerine}{rgb}{1,0.5,0}
\definecolor{Strawberry}{rgb}{1,0,0.5}
\definecolor{Fern}{rgb}{0.25,0.5,0}
\definecolor{Aqua}{rgb}{0,0.5,1}
\definecolor{Moss}{rgb}{0,0.5,0.25}
\definecolor{Mocha}{rgb}{0.5,0.25,0}
\definecolor{Lemon}{rgb}{1,1,0}
\definecolor{Asparagus}{rgb}{0.5,0.5,0}
\definecolor{Grape}{rgb}{0.5,0,1}
\definecolor{Iron}{rgb}{.3,.3,.3}
\definecolor{Steel}{rgb}{.4,.4,.4}
\definecolor{Purple}{rgb}{.5,0,.5}

\def\llbracket{{[\![}}
\def\rrbracket{{]\!]}}
\usepackage[utf8]{inputenc}
\usepackage[table]{xcolor}\usepackage{soul}
\usepackage{afterpage} 

\usepackage{multirow}

\usepackage{makecell} 

\usepackage{tikz}
\usetikzlibrary{backgrounds}
\usepackage[only,llbracket,rrbracket]{stmaryrd}

\usepackage{enumitem} 
\usepackage{color}
\usepackage{array}
\newcolumntype{M}{>{$}c<{$}}

\definecolor{cardinal}{rgb}{0.827, 0, 0}

\begin{document}

\fi

\vfuzz2pt 

\title{
Fire and ice: Partially fault-tolerant quantum computing with selective state filtering
}

\author{Ben W. Reichardt}

\author{David Aasen}

\author{Rui Chao}
\affiliation{Microsoft Quantum}

\begin{abstract}
We develop an error-corrected quantum computation scheme based on concatenating the five-qubit Laflamme code onto the four-qubit Iceberg code.  The approach skates a thin line: it is explicitly not fault tolerant, risking higher logical error rates, and it relies on selective filtering to prepare encoded states for error correction, risking significant overhead.  Yet, at realistic simulated noise rates, the scheme is reliable and resource efficient.  It forges a practical path toward scalable quantum computation.  
\end{abstract}

\maketitle

\section{Introduction}

The next generation of quantum computers is expected to feature two-qubit gate error rates between $10^{-4}$ and $4 \times 10^{-3}$, with trapped ion qubits optimistically nearer the bottom of this range~\cite{Hughes25oxfordionics9999} and other qubit technologies above around $10^{-3}$.  Useful applications will require logical error rates below $10^{-4}$, certainly, and perhaps below $10^{-6}$.  
Pushing for $10^{-9}$, however, may currently be unnecessary given the limited qubit counts and slower speeds of near-term atomic qubit hardware.  This raises a critical question: which error-correcting codes can bridge this gap with minimal space and time overhead?  

In this range of physical and desired logical error rates, it is not necessary to draw from asymptotically large code families like the surface code or more efficient low-density parity check codes.  For platforms with nonlocal qubit connectivity, we argue that small codes (distance $d \leq 8$) are sufficient.  Such codes can offer low qubit overhead and, crucially, simple logical gate implementations.  

In particular, good small codes can be constructed by concatenating onto the $\llbracket 4,2,2 \rrbracket$ ``Iceberg" error-detecting code~\cite{Knill05, BravyiLeemhuisTerhal10majorana, NakaiGoto25iceberg4, BerthusenDursoSabina25iceberg4}.  For example, using the Laflamme $\llbracket 5,1,3 \rrbracket$ or $\llbracket 3,1,2 \rrbracket_4^{\circ 2} = \llbracket 9,1,4 \rrbracket_4$ code as the outer code gives $\llbracket 20,2,6 \rrbracket$ and $\llbracket 36,2,8 \rrbracket$ Calderbank-Shor-Steane (CSS) codes, respectively.  We show that the concatenated structure, together with a ``paired support" property of the outer code (\defref{t:pairedsupport} below), enables efficient state preparation circuits (Figs.~\ref{f:c2026_prep}--\ref{f:c3628_c4848_prep}).  While these circuits are not strictly fault tolerant, they perform very well for Steane-style error correction at moderate error rates.  

For example, in a simplified circuit noise model (\defref{t:noise}) with CNOT depolarization and measurement error rates of $p = 10^{-3}$, our simulations have logical error rates per round of transversal CNOT gates and error correction approximately $2 \times 10^{-6}$ for the $\llbracket 20,2,6 \rrbracket$ code, and $10^{-7}$ for the $\llbracket 36,2,8 \rrbracket$ code.  \tabref{f:simulationssummary} gives error and rejection rates, with $95\%$ confidence intervals, for repeated error correction on one code block, and for transversal CNOTs plus error correction across two blocks.  

\begin{table}
\caption{\label{f:simulationssummary}
For simulations at $p = 10^{-3}$, total ($X$ and $Z$) logical error rates and rejection rates per round of error correction, and per round of transversal CNOT gates and error correction.  
}
\setlength{\tabcolsep}{1.9pt}
\begin{tabular}{rrcc}
\Xhline{2\arrayrulewidth}
& & $\llbracket 20,2,6 \rrbracket$ code & $\llbracket 36,2,8 \rrbracket$ code \\
\hline
\multirow{2}{1.8cm}{\bf Error \\ correction} & Error rate & $1.10(7) \times 10^{-6}$ & \,$5.2^{+1.1}_{-0.9} \times 10^{-8}$ \\
 & Rejection rate & $8.85(4) \times 10^{-5}$ & $4.88(7) \times 10^{-6}$ \\
 \hline
\multirow{2}{1.8cm}{\bf CNOTs} & Error rate & \;\,$2.06(9) \times 10^{-6}$ & \;\;$1.2(1) \times 10^{-7}$ \\
 & Rejection rate & $1.749(8) \times 10^{-4}$ & $1.35(1) \times 10^{-5}$ \\
\Xhline{2\arrayrulewidth}
\end{tabular}
\end{table}

The data in \tabref{f:simulationssummary} need context.  First, the simulations use a simplified noise model; to approximate experimental reality, the CNOT error rate~$p$ must be scaled upward to account for omitted error sources.  
Second, for these even-distance codes, Steane-style error correction rejects ambiguous code blocks (those equidistant from inequivalent codewords).  Because any rejection requires restarting the whole computation, the rejection rate must remain low.  
As argued in \secref{s:postselection}, if the rejection-to-error ratio stays below $300$, for many logical circuits at most $95\%$ of trials will be rejected, leading to a $10 \times$ time overhead; at $p = 10^{-3}$, both codes fall well within this bound.  
Third, our Steane-style error correction and teleportation-based logical gates rely on prepared ancilla states that are tested before use.  If a test fails, that state cannot be used.  Enough states need to be prepared and tested, in parallel or sequentially, so that they are ready to interact with the encoded data, and this requires adaptive routing of ancilla states to data code blocks.  Adaptive routing is also required for magic state distillation, so it is not a new requirement, but it does deserve emphasis.  

The codes presented here---including $\llbracket 20,2,6 \rrbracket$, $\llbracket 32,4,6 \rrbracket$, $\llbracket 36,2,8 \rrbracket$ and $\llbracket 48,4,8 \rrbracket$ codes---demonstrate that leveraging mobile atomic qubits and sacrificing strict fault tolerance can yield high-performance results.  These codes are attractive, resource-efficient candidates for implementing moderate-scale quantum computation on near-term hardware.

\section{Concatenation onto the $\llbracket 4,2,2 \rrbracket$ code}

Superconducting qubits possess limited baked-in connectivity, which restricts the codes that can be used.  The surface code is the benchmark choice, despite its low encoding efficiency, and has already been experimentally demonstrated~\cite{google23surface, google24surfacecode}.  Floquet~\cite{HastingsHaah21floquet} and bivariate bicycle codes~\cite{Yoder25grosscodes} are potential alternatives.  

For mobile qubits like trapped ions or neutral atoms, however, the choice of codes is less constrained by geometry.  Platforms support all-to-all connectivity by moving atoms (at a cost).  
Many small- to medium-sized stabilizer codes with seemingly favorable parameters $\llbracket n,k,d \rrbracket$ are available~\cite{Grassl07codetable}.  This terrain narrows sharply from the requirement that a viable code should support a large set of fault-tolerant logical operations that perform well at realistic noise rates.  Unfortunately, this condition is difficult to verify, so finding good small codes is a challenge.  Simpler properties include CSS structure, low-weight stabilizer generators (or gauge operator checks for subsystem codes), and performant repeated error correction.  

Quantum BCH codes provide a sizeable family of CSS codes~\cite{GrasslBeth99bchcodes}.  This includes promising options, but they generally have high-weight stabilizer generators.  Efficient state preparation and error correction are challenging for the higher-distance BCH codes.  

Instead, we look to code concatenation to build larger codes from very small ones.  In particular, we follow Knill~\cite{Knill05}, who observed that since the $\llbracket 4,2,2 \rrbracket$ code of \figref{f:icebergcode}(a) has two logical qubits, we can naturally concatenate onto it an $\llbracket n,k,d \rrbracket_4$ ``ququad" code.  (The subscripted $4$ indicates a stabilizer code in which the subsystems are pairs of qubits, with four dimensions, rather than qubits; thus $2n$ physical qubits encode $2k$ logical qubits.)  The concatenated code, $\llbracket n,k,d \rrbracket_4 \circ \llbracket 4,2,2 \rrbracket$, is a $\llbracket 4n, 2k, 2d \rrbracket$ qubit code.  Knill's scheme uses the $\llbracket 3,1,2 \rrbracket_4$ self-dual CSS code of \figref{f:icebergcode}(b) concatenated $\ell$ times, to get $\llbracket 3,1,2 \rrbracket_4^{\circ \ell} \circ \llbracket 4,2,2 \rrbracket$ codes.  Repeated error correction with the first of these, the $\llbracket 12,2,4 \rrbracket$ code of \figref{f:icebergcode}(c), has been implemented in experiment~\cite{Silva24microsoft12qubitcode}.  

\begin{figure}
\centering
\begin{minipage}[b]{0.2\textwidth}
\centering
$$\begin{array}{r c c c c}
&X&X&X&X\\
&Z&Z&Z&Z\\ \cline{2-5}
X_{1,L}\!:&X&X&I&I\\
Z_{1,L}\!:&I&Z&I&Z\\
X_{2,L}\!:&X&I&X&I\\
Z_{2,L}\!:&I&I&Z&Z
\end{array}$$
(a)
\end{minipage}
\hfill 
\begin{minipage}[b]{0.2\textwidth}
\centering
$$\begin{array}{r c c c}
&XI&IX&XX\\
&XX&XI&IX\\
&ZI&IZ&ZZ\\
&ZZ&ZI&IZ\\ \cline{2-4}
X_{1,L}\!:&XX&II&XI\\
Z_{1,L}\!:&ZZ&II&ZI\\
X_{2,L}\!:&IX&II&XX\\
Z_{2,L}\!:&IZ&II&ZZ
\end{array}$$
(b) \label{}
\end{minipage}
\hfill
\begin{minipage}[b]{0.5\textwidth}
\centering
$$
\def\I{\cdot}
\begin{array}{r c @{\;} c @{\;} c @{\;} c @{\;\;\;} c @{\;} c @{\;} c @{\;} c @{\;\;\;} c @{\;} c @{\;} c @{\;} c}
&X&X&X&X&\I&\I&\I&\I&\I&\I&\I&\I\\
&Z&Z&Z&Z&\I&\I&\I&\I&\I&\I&\I&\I\\
&\I&\I&\I&\I&X&X&X&X&\I&\I&\I&\I\\
&\I&\I&\I&\I&Z&Z&Z&Z&\I&\I&\I&\I\\
&\I&\I&\I&\I&\I&\I&\I&\I&X&X&X&X\\
&\I&\I&\I&\I&\I&\I&\I&\I&Z&Z&Z&Z\\
&X&X&\I&\I&X&\I&X&\I&\I&X&X&\I\\
&\I&X&X&\I&X&X&\I&\I&X&\I&X&\I\\
&\I&Z&\I&Z&\I&\I&Z&Z&\I&Z&Z&\I\\
&\I&Z&Z&\I&\I&Z&\I&Z&\I&\I&Z&Z\\ \cline{2-13}
X_{1,L}\!:&\I&X&X&\I&\I&\I&\I&\I&X&X&\I&\I\\
Z_{1,L}\!:&\I&Z&Z&\I&\I&\I&\I&\I&\I&Z&\I&Z\\
X_{2,L}\!:&X&\I&X&\I&\I&\I&\I&\I&\I&X&X&\I\\
Z_{2,L}\!:&\I&\I&Z&Z&\I&\I&\I&\I&\I&Z&Z&\I
\end{array}$$
(c) \label{}
\end{minipage}
\caption{
(a) The $\llbracket 4,2,2 \rrbracket$ code encodes two logical qubits in four physical qubits, protected to distance two, i.e., detecting single-qubit errors.  
(b) The $\llbracket 3,1,2 \rrbracket_4$ code encodes two logical qubits in six physical qubits, detecting one- and two-qubit errors on any of the three qubit pairs.  
(c) The concatenation $\llbracket 3,1,2 \rrbracket_4 \circ \llbracket 4,2,2 \rrbracket$ gives a $\llbracket 12,2,4 \rrbracket$ code.  We have used $\cdot$ for trivial entries to bring out the structure.  
} \label{f:icebergcode}
\end{figure}

An $\llbracket n,k,d \rrbracket_4$ CSS ququad code can be made from two copies of an $\llbracket n,k,d \rrbracket$ CSS qubit code.  
The many-hypercube codes~\cite{Goto24manyhypercube, NakaiGoto25iceberg4}, for example, are defined by concatenating the $\llbracket 4,2,2 \rrbracket$ code with itself in this way.  
More interestingly, 

\begin{lemma}[\cite{BravyiLeemhuisTerhal10majorana}] \label{t:BravyiLeemhuisTerhal10majorana}
Any $\llbracket n,k,d \rrbracket$ stabilizer code $\cC$ can be mapped to an $\llbracket n,k,d \rrbracket_4$ CSS code $\D(\cC)$.  
\end{lemma}

\noindent
Indeed, a stabilizer $P = \pm \bigotimes_{j = 1}^n P_j \in \pm \{I, X, Y, Z\}^{\otimes n}$ of $\cC$ gives an $X$ stabilizer $\D_X(P)$ of the new code by applying $\D_X: I \mapsto II, X \mapsto XI, Z \mapsto IX, Y \mapsto XX$ to each $P_j$, and gives a $Z$ stabilizer $\D_Z(P)$ by applying $\D_Z: I \mapsto II, X \mapsto IZ, Z \mapsto ZI, Y \mapsto ZZ$ to each $P_j$.  

For example, the $\llbracket 5,1,3 \rrbracket$ code of \figref{f:2026code}(a) maps to the ten-qubit, $\llbracket 5,1,3 \rrbracket_4$ CSS code of \figref{f:2026code}(b).  

\begin{figure}
\centering
\begin{minipage}[b]{0.15\textwidth}
\centering
\def\I{\cdot}
\raisebox{2.35cm}{$
\begin{array}{r @{\;} c @{\;} c @{\;} c @{\;} c @{\;} c}
&X&Z&Z&X&\I\\
&Z&Y&Y&Z&\I\\
&\I&X&Z&Z&X\\
&\I&Z&Y&Y&Z\\ \cline{2-6}
X_L\!:&X&Y&X&\I&\I\\
Z_L\!:&Z&X&Z&\I&\I
\end{array}
$}
(a)
\end{minipage}
\begin{minipage}[b]{0.3\textwidth}
\centering
$$\def\I{\cdot}
\overset{\text{\lemref{t:BravyiLeemhuisTerhal10majorana}}}{\longmapsto}
\!\!\!\!\!\!\!\begin{array}{r c c c c c}
&XI&IX&IX&XI&\I\\
&IX&XX&XX&IX&\I\\
&\I&XI&IX&IX&XI\\
&\I&IX&XX&XX&IX\\
&IZ&ZI&ZI&IZ&\I\\
&ZI&ZZ&ZZ&ZI&\I\\
&\I&IZ&ZI&ZI&IZ\\
&\I&ZI&ZZ&ZZ&ZI \\ \cline{2-6}
X_{1,L}\!:&XI&XX&XI&\I&\I \\
Z_{1,L}\!:&ZI&IZ&ZI&\I&\I \\
X_{2,L}\!:&IX&XI&IX&\I&\I \\
Z_{2,L}\!:&IZ&ZZ&IZ&\I&\I
\end{array}$$
(b) \label{}
\end{minipage} \\
\begin{minipage}[b]{0.5\textwidth}
\centering
$$\def\I{\cdot}
\overset{\text{\corref{t:ququadiceberg}}}{\longmapsto}
\!\!\!\!\!\!\!\begin{array}{r c@{\,}c@{\,}c@{\,}c c@{\,}c@{\,}c@{\,}c c@{\,}c@{\,}c@{\,}c c@{\,}c@{\,}c@{\,}c c@{\,}c@{\,}c@{\,}c}
&1&1&1&1&\I&\I&\I&\I&\I&\I&\I&\I&\I&\I&\I&\I&\I&\I&\I&\I\\
&\I&\I&\I&\I&1&1&1&1&\I&\I&\I&\I&\I&\I&\I&\I&\I&\I&\I&\I\\
&\I&\I&\I&\I&\I&\I&\I&\I&1&1&1&1&\I&\I&\I&\I&\I&\I&\I&\I\\
&\I&\I&\I&\I&\I&\I&\I&\I&\I&\I&\I&\I&1&1&1&1&\I&\I&\I&\I\\
&\I&\I&\I&\I&\I&\I&\I&\I&\I&\I&\I&\I&\I&\I&\I&\I&1&1&1&1\\
&1&1&\I&\I&1&\I&1&\I&1&\I&1&\I&1&1&\I&\I&\I&\I&\I&\I\\
&1&\I&1&\I&\I&1&1&\I&\I&1&1&\I&1&\I&1&\I&\I&\I&\I&\I\\
&\I&\I&\I&\I&1&1&\I&\I&1&\I&1&\I&1&\I&1&\I&1&1&\I&\I\\
&\I&\I&\I&\I&1&\I&1&\I&\I&1&1&\I&\I&1&1&\I&1&\I&1&\I\\ \cline{2-21}
X_{1,L}, Z_{2,L}:&1&1&\I&\I&\I&1&1&\I&1&1&\I&\I&\I&\I&\I&\I&\I&\I&\I&\I\\
X_{2,L}, Z_{1,L}:&1&\I&1&\I&1&1&\I&\I&1&\I&1&\I&\I&\I&\I&\I&\I&\I&\I&\I
\end{array}$$
(c) \label{}
\end{minipage}
\caption{
(a) Stabilizer generators and logical operators for a $\llbracket 5,1,3 \rrbracket$ code.  
(b) Applying \lemref{t:BravyiLeemhuisTerhal10majorana} gives a $\llbracket 5,1,3 \rrbracket_4$ CSS code.  
(c) Concatenating with the $\llbracket 4,2,2 \rrbracket$ code gives a $\llbracket 20,2,6 \rrbracket$ self-dual CSS code.  The $X$ and $Z$ operators have the same supports, indicated with $1$s.  
} \label{f:2026code}
\end{figure}

\begin{corollary}[\cite{BravyiLeemhuisTerhal10majorana}] \label{t:ququadiceberg}
An $\llbracket n,k,d \rrbracket$ stabilizer code can, after the mapping of \lemref{t:BravyiLeemhuisTerhal10majorana}, be concatenated with the $\llbracket 4,2,2 \rrbracket$ code, yielding a $\llbracket 4n, 2k, 2d \rrbracket$ self-dual CSS code.  
\end{corollary}

\noindent (While we have credited \cite{BravyiLeemhuisTerhal10majorana} for these claims, the same observation was made in the qubit formalism by~\cite{KovalevPryadko13symplecticdouble}, and \cite{Burton24symplecticdouble} have more recently phrased it in terms of concatenation with the $\llbracket 4,2,2 \rrbracket$ code.)  

Figure~\ref{f:2026code}(c) shows the $\llbracket 20,2,6 \rrbracket$ code that results from applying the corollary to concatenate $\llbracket 5,1,3 \rrbracket \circ \llbracket 4,2,2 \rrbracket$.  

\medskip

Berthusen and Durso-Sabina~\cite{BerthusenDursoSabina25iceberg4} have independently recently studied fault-tolerance schemes based on using \lemref{t:BravyiLeemhuisTerhal10majorana} to concatenate onto the $\llbracket 4,2,2 \rrbracket$ code.  They consider some of the same codes, with similar assumptions, but with very different, less efficient but fault-tolerant, state-preparation circuits.  They use a more complicated decoder, relying on belief propagation rather than a lookup table.  They find similar logical error rates to us for repeated Steane-style error correction with the $\llbracket 20,2,6 \rrbracket$ code, and without any rejection (postselection).  
However, they do not use full circuit-level noise, but fit a homogeneous independent noise model to the state preparation subroutine~\cite{Eastin06homogeneous}.  
They also do not simulate logical operations.  We find that transversal CNOT gates, with error correction, can have very different logical error rates than repeated error correction alone.

\section{The role of postselection} \label{s:postselection}

Postselection is a technique in which we discard a trial when the syndrome data indicates that a logical error is more likely to have occurred, because a state is far from the codespace.  Discarding the least reliable trials reduces the logical error rate, but the tradeoff is that more trials are needed.  Important quantum subroutines, especially magic state distillation and cultivation~\cite{GidneyShuttyJones24cultivation}, are known to benefit greatly from selecting on the state being in or near the codespace.  How generic is this improvement, and what will be the role of postselection in future fault-tolerant quantum algorithms?  In particular, for a code with even distance~$d$, the per-operation probability of a logical error, $p_L$, will scale like $c\, p^{d/2+1}$, where $p$ is a small-enough physical error rate (possibly scaling multiple parameters of a complicated physical model) and $c$ is a code-dependent constant, because $d/2+1$ faults can be decoded in the wrong direction.  The probability of rejecting due to an ambiguous syndrome, $p_R$, will scale like $c' p^{d/2}$ for a different constant $c'$.  

As a concrete example, for the $\llbracket 36,2,8 \rrbracket$ code described below, under independent bit-flip memory noise at rate~$p$, we find $c = 54432$ (the number of subsets of $5$ qubits on which bit flips decode to a logical error) and $c' = 23544$ (the number of subsets of $4$ qubits for which bit flips cannot be decoded).  

Consider a routine with $n$ operations satisfying 
\begin{enumerate}
\item The total logical error rate is at most $p_L^* = 1\%$ 
\item The acceptance rate is at least $1 - p_R^* = 5\%$
\end{enumerate}
Let us make a back-of-the-envelope calculation, assuming independent failures.  The first condition requires $n \, p_L \leq p_L^*$, limiting $n \leq p_L^* / p_L$.  
The second condition also limits the number of operations.  The probability of accepting is $(1 - p_R)^n < e^{- p_R n}$.  For this to be at least $1 - p_R^*$, $n \leq -\log(1 - p_R^*) / p_R$.  
Thus, whenever $p_L^* / p_L < -\log(1 - p_R^*) / p_R$, or $p_R / p_L < -\log(1 - p_R^*) / p_L^* \approx 300.$, $n$ is limited by the logical error rate and not the acceptance rate.  In our simulations, we will generally find that $p_R / p_L < 300$ for realistic values of~$p$, so $n$ is limited by $p_L$.  For very small~$p$, however, namely $p < c' / (300 c)$, $p_R / p_L > 300$, so the rejection rate limits the number of possible operations.  

A $5\%$ acceptance rate implies a $10 \times$ time overhead---not $20 \times$, since rejections are on averaged detected halfway through.  Whether this overhead is acceptable is debatable and depends on one's computational budget.  If we require a $10\%$ acceptance rate, then the critical value for $p_R / p_L$ decreases to $230.$, while if we can tolerate a $1\%$ acceptance rate, the critical value becomes about $460$.  Perhaps more interesting is the restriction on the total logical error rate.  Problems with a verifiable solution, for example factoring, are interesting even with much higher logical error rates.  Most smaller-scale quantum algorithms do not have easily verifiable answers, unfortunately, so it is important to keep the logical error rate low.  For example, for a preparing a state to within ``chemical accuracy," we want the total logical error rate to be less than $0.1\%$.  If a $10\%$ logical error rate is acceptable, though, then postselection limits the number of operations that we can run whenever $p_R / p_L > 30$, a much larger range.

\section{Steane-style error correction}

\begin{figure}
\includegraphics[scale=1]{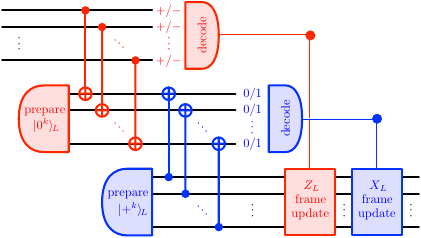}
\caption{Steane-style $Z$ error correction (red) and $X$ error correction (blue) for an $\llbracket n,k,d \rrbracket$ CSS code.  Since transversal physical CNOTs implement transversal logical CNOTs, the logical circuit is two steps of one-qubit teleportation.  The difficult part of implementing this scheme is constructing good circuits to prepare the encoded ancilla states $\ket{0^k}_{\!L}$ and~$\ket{+^k}_{\!L}$.  
} \label{f:steanestyleec}
\end{figure}

In Steane-style error correction for an $\llbracket n,k,d \rrbracket$ CSS code~\cite{Steane97}, $Z$ and $X$ error correction are implemented using encoded $\ket{0^k}$ and encoded $\ket{+^k}$ states, transversal CNOT gates, and transversal measurements.  We use the variant in \figref{f:steanestyleec}~\cite{Silva24microsoft12qubitcode} where the logical data is teleported into fresh ancilla blocks twice; this refreshes the data qubits, automatically correcting for qubit leakage and loss errors.  The main challenge is to construct reliable, resource-efficient circuits for encoded $\ket{0^k}$ and $\ket{+^k}$.  For the moderate-size codes that we consider, decoding can be performed efficiently using a small precomputed lookup table that maps syndromes to corrections.  

One can develop a useful intuitive model for repeated Steane-style error correction by assuming that the $\ket{0^k}_L$ and $\ket{+^k}_L$ ancilla states have independent, identically distributed errors on each qubit~\cite{Eastin06homogeneous}.   
Then in steady state the relevant qubit error rate for the $X$ error decoder is roughly $p_X(\ket{0^k}_L) + 2 p_X(\ket{+^k}_L) + 3 p_X(\text{CNOT}) + p_X(\text{measurement})$, where $p_X(\ket{\psi}_L)$ is the $X$ error rate on that state and $p_X(\text{CNOT})$ is the marginal probability of an $X$ error on a qubit after a CNOT gate.  Good preparation circuits should keep $p_X(\ket{\psi}_L)$ not much larger than $p_X(\text{CNOT})$.  After several crude assumptions, we thus intuit that the logical $X$ error rate introduced by one full error-correction round should be roughly the code's logical error rate for physical memory errors at rate $c \, p_X(\text{CNOT})$ for a small constant~$c$.  

(Knill-style error correction, based on logical teleportation through encoded Bell pairs, should be able to reach lower logical error rates.  A similar argument gives that the relevant error rate for the $X$ error decoder is only roughly $p_X\big((\ket{00}+\ket{11})^{\otimes k}\big) + p_X(\text{CNOT}) + p_X(\text{measurement})$.  We have focused on Steane-style error correction to avoid excessive preselection overhead.)  

While the assumptions are wrong, the above conclusion is supported by our simulations.  The logical error rate of repeated Steane-style error correction with circuit-level noise is approximately predicted by the logical error rate of a bit-flip ($X$ error) noise channel with an ideal decoder.  Let us therefore start by studying code performance under bit-flip memory noise.

\section{Code performance under bit-flip memory noise}

\begin{figure*}
\includegraphics[scale=.42]{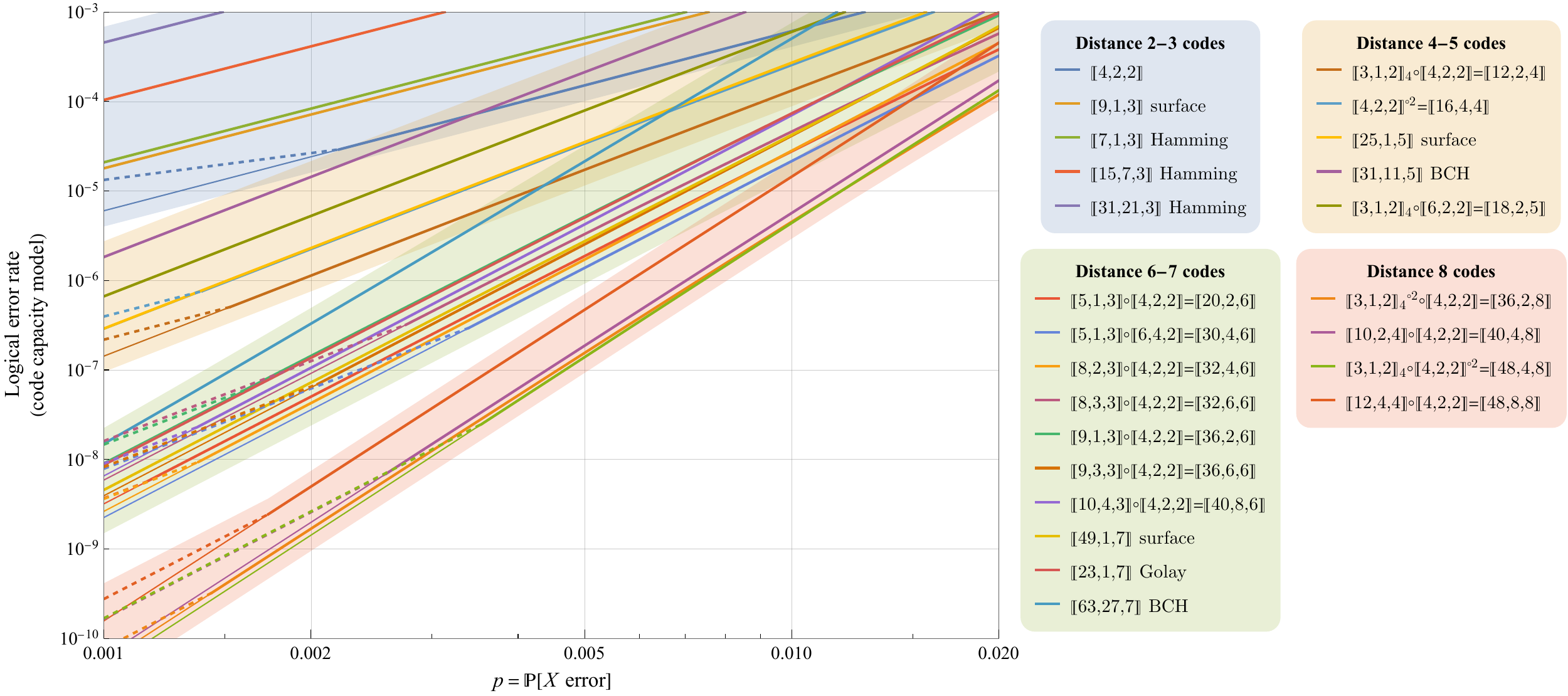}
\caption{Logical error rate conditioned on acceptance, $p_L$, for an ideal decoder against independent bit-flip noise at rate~$p$, i.e., code capacity model.  The dashed lines plot $p_R / 300$ in the ranges when the rejection rate limits the number of operations according to the criteria of \secref{s:postselection}, i.e., when $p_R > 300 p_L$.  
} \label{f:memorynoiseplot}
\end{figure*}

\begin{figure}
\includegraphics[scale=.405]{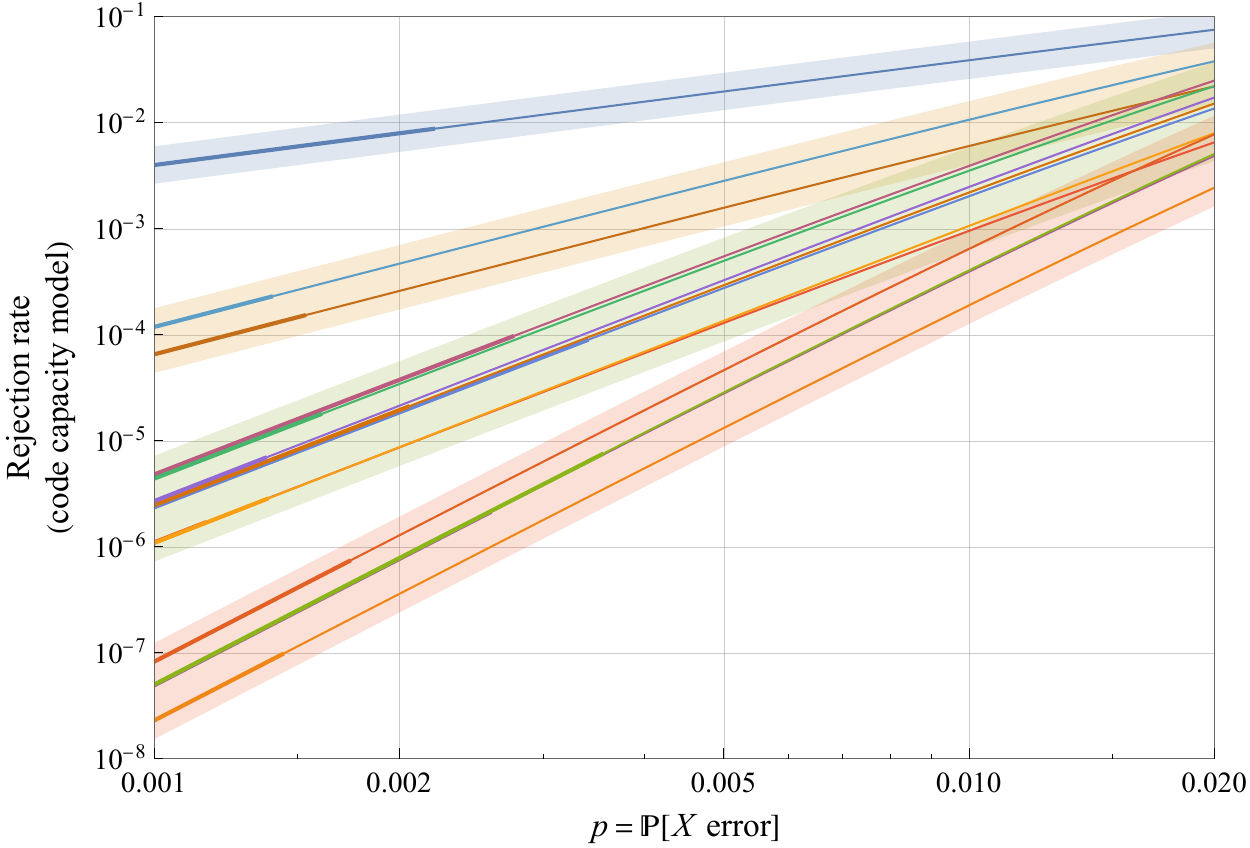}
\caption{Rejection rate $p_R$ for an ideal decoder against independent bit-flip noise at rate~$p$.  The legend is the same as in \figref{f:memorynoiseplot} except not showing the odd-distance codes without postselection.  The curves are thicker when $p_R > 300 p_L$.} \label{f:memorynoiseplotrejection}
\end{figure}

For a variety of promising codes, we compute the logical error rate of an ideal decoder after independent bit-flip $X$ noise at rate~$p$, i.e., the code capacity model.  Figures~\ref{f:memorynoiseplot} and~\ref{f:memorynoiseplotrejection} plot $p_L$ and $p_R$ for the different codes.  
The codes we consider include both small codes concatenated onto the $\llbracket 4,2,2 \rrbracket$ code, and other common CSS codes as well for context~\cite{AlbertFaist25errorcorrectionzoo}.  

Although we argued above that Steane-style error correction, with circuit-level noise, can be approximately modeled with the code capacity model, this remains a significant simplification.  Furthermore, for the surface codes, the usual, geometrically local, error-correction method is ``Shor-style," based on repeatedly measuring stabilizers, rather than Steane-style.  
We can still, however, extract useful lessons from these figures.  

\begin{enumerate}[leftmargin=*]
\item 
For low enough~$p$, the higher-distance codes generally have lower logical error rates than lower-distance codes, a threshold behavior.  
\item
Codes in the same distance range, $d \in \{2t, 2t+1\}$, generally have similar logical error rates~$p_L$.  A simple method for choosing a code might start by choosing the desired distance, then choosing the best code offering that distance protection.  
\item 
The codes with even distance $2t$, codes that use post\-selection, generally perform comparably or even better than the odd-distance $2t+1$ codes with no postselection.  In both cases, it takes at least $t+1$ $X$ faults to cause a logical error.  For small~$p$, therefore, the slope of the $p_L$ curve on a log-log scale approaches $t+1$.  
\item 
For most of the codes, the rejection rate is not the limiting factor across most of the range, i.e., $p_R < 300 p_L$, for $p \gtrsim 0.003$.  
\item
For the same $t$, codes with more logical qubits $k$ per code block usually also have higher logical error rates $p_L$.  Intuitively, this is because the same distance protection is spread more thinly over more physical qubits.  The number of weight $t+1$ errors that cause a logical error can grow as $\binomial{n}{t+1}$, which grows with $n$ for a fixed~$t$.  

Note that had we plotted $p_L / k$ instead of $p_L$, the codes with the same distance would be even more tightly clustered.  Whether such normalization is appropriate depends on the ultimate application and the ease or difficult of implementing logical operations with a code.  For pure memory, i.e., repeated error correction, normalizing by $k$ makes sense.  For computation, it is less clear, since codes with larger $k$ often require more complicated ways to address different logical qubits.  
\end{enumerate}

Note that there are at least two ways of concatenating an $\llbracket n,k,d \rrbracket$ CSS code onto the $\llbracket 4,2,2 \rrbracket$ code to obtain a CSS code.  One method is through \corref{t:ququadiceberg}, yielding a $\llbracket 4n,2k,2d \rrbracket$ self-dual CSS code.  Another method is to take $n$ $\llbracket 4,2,2 \rrbracket$ code blocks, and to impose the outer code's stabilizers across all $n$ of the first logical qubits, and separately across all $n$ second logical qubits.  This gives a $\llbracket 4n,2k,2d \rrbracket$ CSS code that is not necessarily self dual.  For the $\llbracket 4,2,2 \rrbracket$ code concatenated on itself, these two concatenation methods are equivalent.  For the $\llbracket 3,1,2 \rrbracket_4 \circ \llbracket 4,2,2 \rrbracket$ code concatenated onto the $\llbracket 4,2,2 \rrbracket$ code, the two methods are not equivalent.  We have used the second method, based on two copies of the outer $\llbracket 12,2,4 \rrbracket$ code, because this turns out to have a simpler encoding circuit.  We will discuss encoding circuits for codes concatenated onto the $\llbracket 4,2,2 \rrbracket$ code in \secref{s:statepreparationcircuits} below.  

A key metric for evaluating a fault-tolerance scheme is its overhead.  Figures~\ref{f:memorynoiseplot} and~\ref{f:memorynoiseplotrejection} provide insight into error rates but reveal nothing about overhead---a critical factor, especially at higher physical error rates~$p$.  While larger-distance codes typically incur greater overhead, this cost is not captured by the inverse rate $n/k$; in fact, $n/k$ often severely underestimates the true overhead.  We will also examine overhead in detail in \secref{s:statepreparationcircuits}.

\section{Repeated error correction simulations}

Figure~\ref{f:memorynoiseplot} suggests that achieving logical error rates per round of error correction of $\lesssim 10^{-6}$, at moderate physical noise rates, will require using codes of distance $\geq 6$.  \appref{s:distance5codes} simulates several codes of distance 2--5 under circuit-level noise, essentially validating this conclusion.  In this section, we will study repeated error correction for codes of distance 6--8.  

We will consider the $\llbracket 20,2,6 \rrbracket$ code of \figref{f:2026code}, the $\llbracket 32,4,6 \rrbracket$ code obtained from applying \corref{t:ququadiceberg} to the $\llbracket 8,2,3 \rrbracket$ code of \figref{f:823code}, the $\llbracket 3,1,2 \rrbracket_4^{\circ 2} \circ \llbracket 4,2,2 \rrbracket = \llbracket 36,2,8 \rrbracket$ Knill code, and the $\big( \llbracket 3,1,2 \rrbracket_4 \circ \llbracket 4,2,2 \rrbracket)^{\otimes 2} \circ \llbracket 4,2,2 \rrbracket = \llbracket 48,4,8 \rrbracket$ code.  Aside from their solid performance under memory bit-flip noise, the key reason for choosing these codes is the ease of preparing encoded $\ket{0^k}$.  We will describe the state-preparation circuits below.  

\begin{figure}
\centering
$$\def\I{\cdot}
\begin{array}{r c c c c c c c c}
&X&X&X&X&\I&\I&\I&\I\\
&Z&Z&Z&Z&\I&\I&\I&\I\\
&\I&\I&\I&\I&X&X&X&X\\
&\I&\I&\I&\I&Z&Z&Z&Z\\
&\I&X&Y&Z&\I&X&Y&Z\\
&\I&Z&X&Y&\I&Z&X&Y\\ \cline{2-9}
X_{1,L}\!:&\I&\I&\I&\I&\I&X&Z&Y\\
Z_{1,L}\!:&\I&\I&\I&\I&\I&Z&Y&X\\
X_{2,L}\!:&\I&X&Z&Y&\I&\I&\I&\I\\
Z_{2,L}\!:&\I&Z&Y&X&\I&\I&\I&\I
\end{array}$$
\caption{
Stabilizer generators and logical operators for an $\llbracket 8,2,3 \rrbracket$ code.  
} \label{f:823code}
\end{figure}

We simulate stochastic Pauli noise with parameter~$p$: 

\begin{definition}[Noise$(p)$] \label{t:noise}
Every two-qubit gate is independently depolarized with probability $\tfrac{16}{15} p$, so $p$ is the probability of a non-trivial Pauli error added after the gate.  Also, measurements are incorrect with probability~$p$.  Qubit preparations and one-qubit gates are perfect.  
\end{definition}

\noindent
For now, we do not simulate noise on idle qubits, any bias in the two-qubit gate errors, nor leakage and loss.  For rough estimates, additional noise sources can be accounted for by increasing~$p$.  The teleportation-based error correction we use naturally corrects leaked or lost qubits.  

For evaluating repeated error correction with circuit-level noise, we simulate state preparation followed by ten rounds of $X$ and $Z$ error correction, followed by transversal $Z$ measurement and decoding: 
\begin{equation}
\raisebox{-.6cm}{
\includegraphics[scale=1]{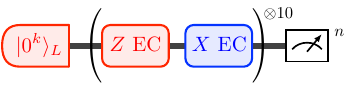}
}
\end{equation}
Each $X$ or $Z$ error correction is decoded independently using the same lookup-table decoder as we used in the code-capacity model calculations of Figs.~\ref{f:memorynoiseplot}--\ref{f:memorynoiseplotrejection}.  
Ten rounds of error correction are more than enough to approach a steady state distribution, away from the transient effects at the beginning.  We divide the resulting rejection rates and logical $X$ error rates by $10$ to get per-round rates.  

In order to put the logical error rates in context, we also simulate the $\llbracket d^2, 1, d \rrbracket$ rotated surface codes~\cite{BombinMartindelgado07surfaceoverhead} for odd $d \leq 9$.  For the surface codes, we simulate $10$ cycles of noisy syndrome extraction, using the PyMatching decoder~\cite{Higgott2022PyMatching}, then divide the logical error rate by $10$.  Arguably, this comparison unfairly advantages the surface codes, since fault-tolerant error correction for the surface codes takes $d$ cycles of syndrome extraction.  In non-local architectures, algorithmic fault tolerance~\cite{Zhou25algorithmicfaulttolerance, Terhal25algorithmicfaulttolerance} might allow reducing the number of syndrome extraction cycles between transversal logical operations.  However, the other codes, too, especially the Iceberg codes, should benefit from allowing multiple logical operations between error correction or detection~\cite{Zalka97}.  

\begin{figure}
\centering
\subfigure[]{
\includegraphics[scale=.41]{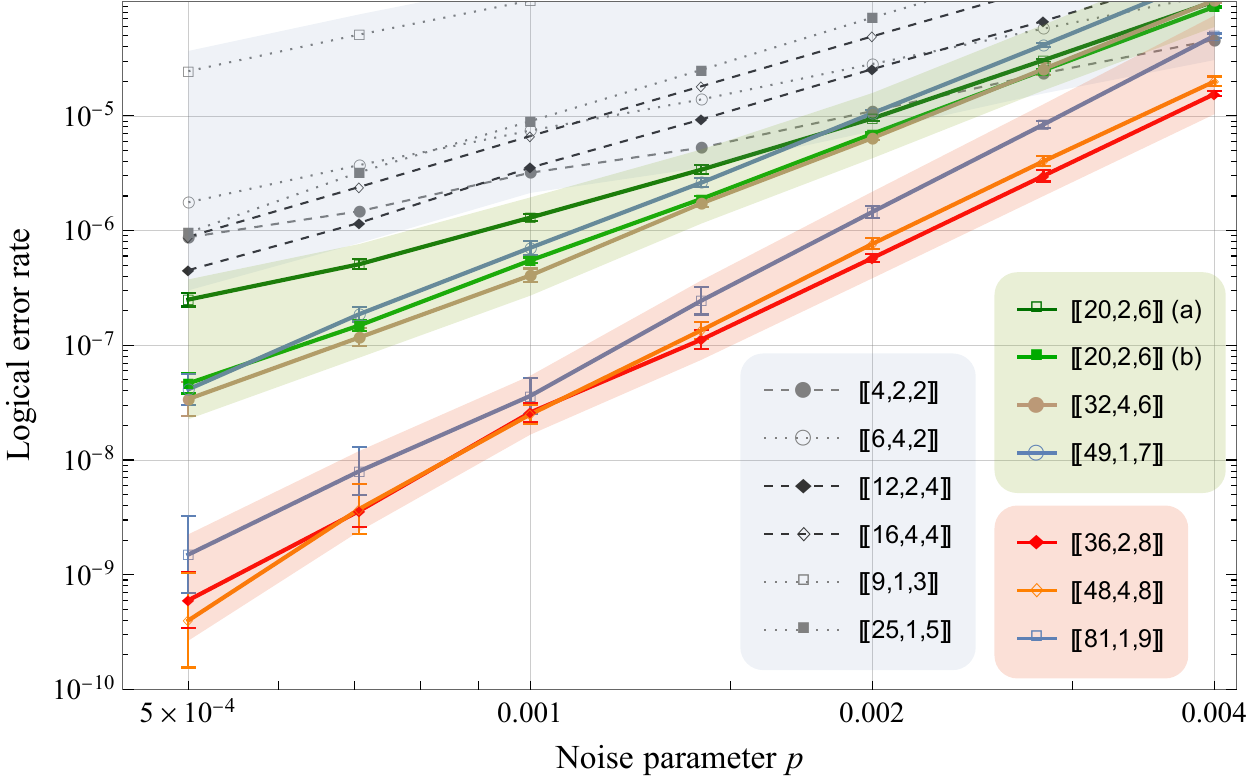}
}
\subfigure[]{
\includegraphics[scale=.41]{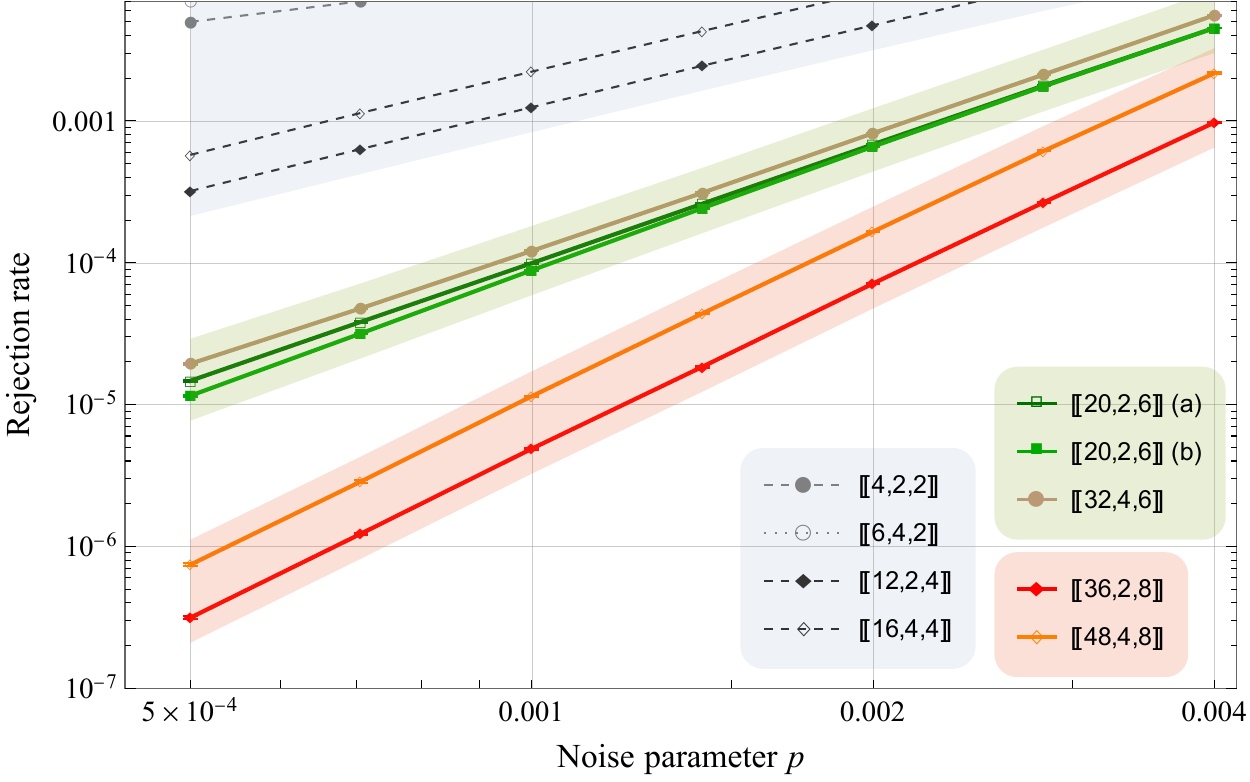}
}
\caption{Logical error rate (a) and rejection rate (b) per round of error correction, or per cycle of syndrome extraction for the surface codes.  The surface codes have rejection rate $0$.  The gray data points, for codes of distance $\leq 5$, are plotted more clearly in \figref{f:lowdistancecodes}.  
} \label{f:logicalerrorplot}
\end{figure}

The simulated logical error and rejection rates are plotted in \figref{f:logicalerrorplot}.  Broadly, the higher-distance codes give lower error and rejection rates.  

\begin{figure}
\centering
\subfigure[\label{f:ancillaacceptanceplot}]{\includegraphics[scale=.41]{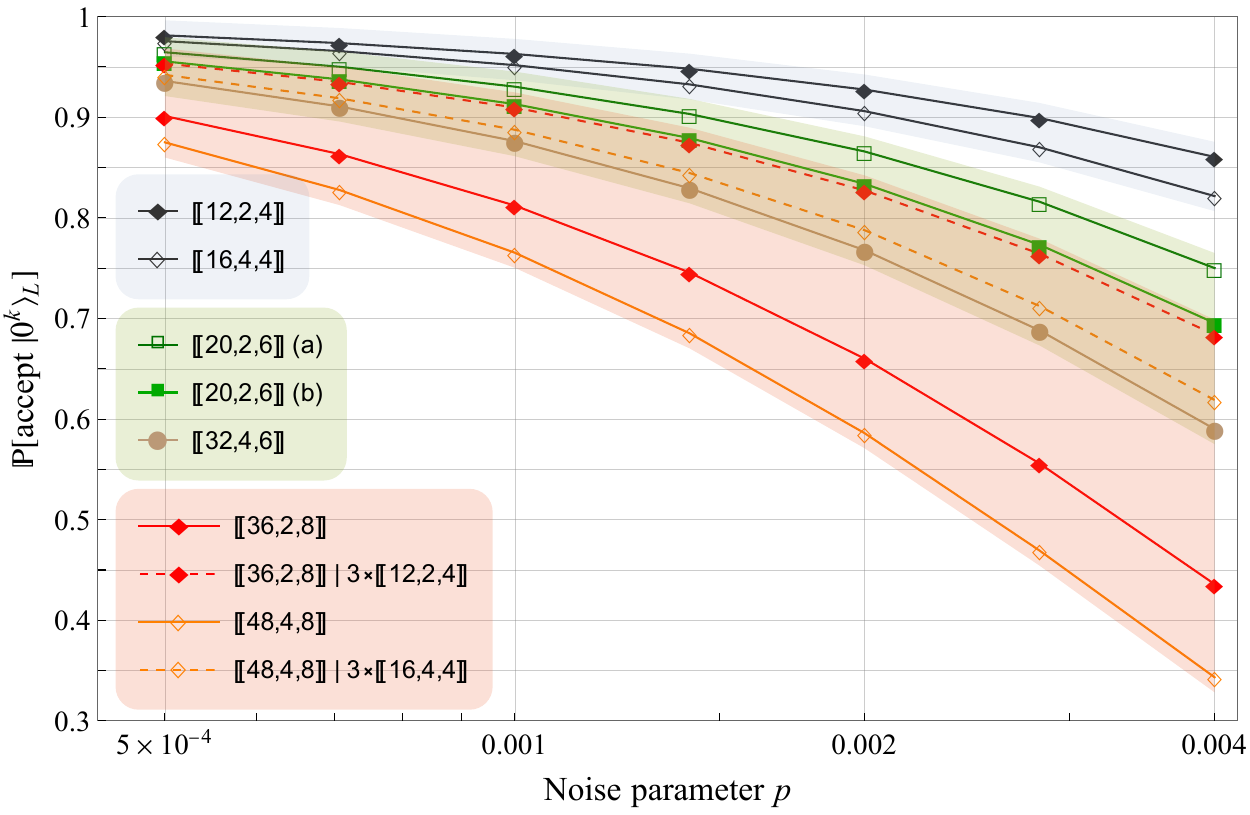}}
\subfigure[\label{f:ecnotsplot}]{\includegraphics[scale=.41]{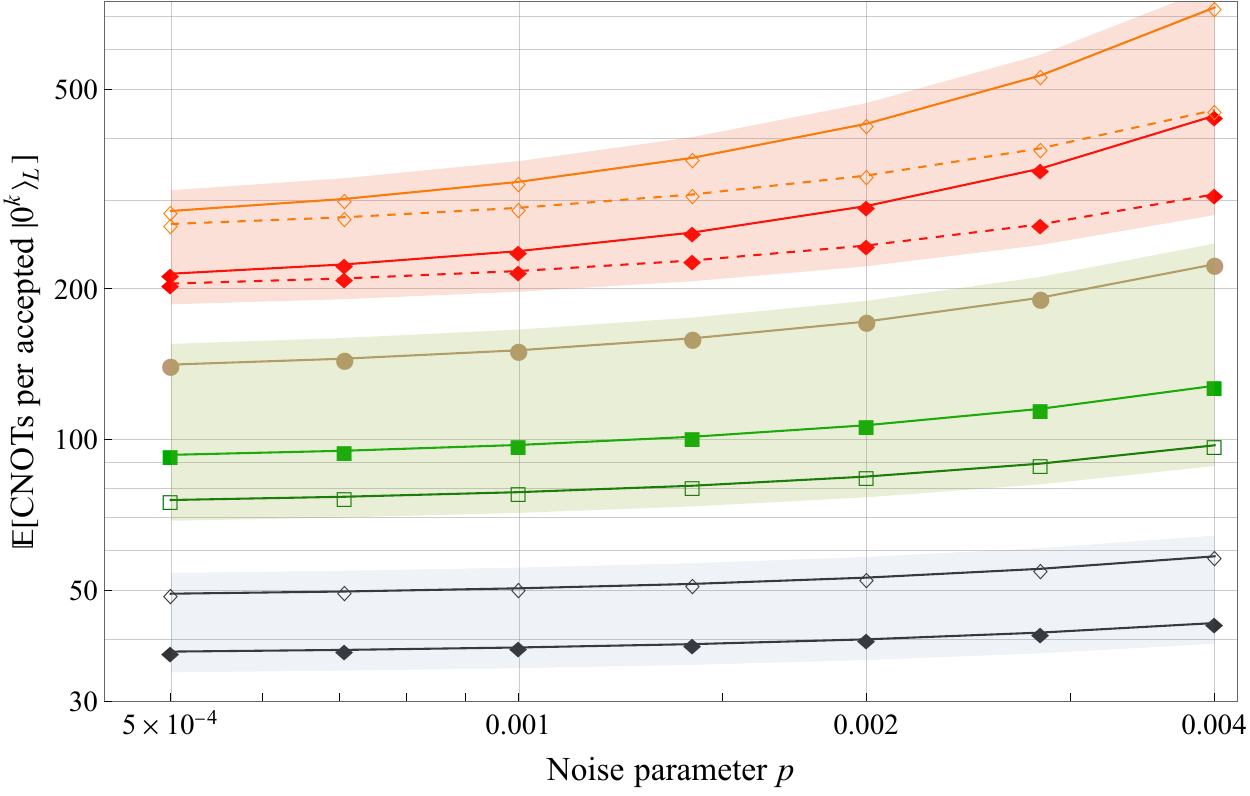}}
\caption{Overhead estimation.  (a) Acceptance probabilities for preparing encoded $\ket{0^k}$.  The two dashed lines give the acceptance probabilities for the $\llbracket 36,2,8 \rrbracket$ and $\llbracket 48,4,8 \rrbracket$ codes conditioned on being given three accepted copies of $\ket{0^k}_L$ for the $\llbracket 12,2,4 \rrbracket$ and $\llbracket 16,4,4 \rrbracket$ codes, respectively.  For these codes, a two-stage ancilla factory can be considerably more efficient than a single-stage factory in which we reject the full state when only a part of it has failed.  
(b) The expected number of CNOT gates used to prepare an accepted $\ket{0^k}_L$ state, with the same legend.  The dashed lines give the expected number of CNOTs assuming a two-stage ancilla factory.  
} \label{f:ancillaacceptance}
\end{figure}

This comes at the cost of higher overhead, however.  The main contributor to the overhead is the cost of preparing the encoded $\ket{0^k}$ and $\ket{+^k}$ ancilla states for Steane-style error correction.  In particular, we reject a prepared state when an error is detected, so multiple states might need to be prepared before one passes and can be used in error correction.  Figure~\ref{f:ancillaacceptanceplot} plots the probability of accepting an encoded ancilla state versus~$p$ for the different codes.  The acceptance probability drops exponentially with~$p$, and more rapidly with larger block codes.  Figure~\ref{f:ecnotsplot} plots the expected number of CNOT gates needed to prepare an accepted ancilla state.  This shows clearly the high overhead cost of using higher-distance codes, especially at larger~$p$.  

In these plots, there are two curves for the $\llbracket 20,2,6 \rrbracket$ code, corresponding to two different preparation circuits that we will describe below.  The second circuit has higher overhead, but performs significantly better at low error rates, $p \lesssim 0.1\%$.  We will explain this performance divergence in \secref{s:statepreparationcircuits} below.  
In \figref{f:ancillaacceptance}, there are also two curves for the $\llbracket 36,2,8 \rrbracket$ and $\llbracket 48,4,8 \rrbracket$ codes.  The lower-overhead dashed curves correspond to using a two-stage ancilla factory that we will also explain below.

\section{State-preparation circuits} \label{s:statepreparationcircuits}

\begin{figure}
\centering
\subfigure[]{\includegraphics[scale=.9]{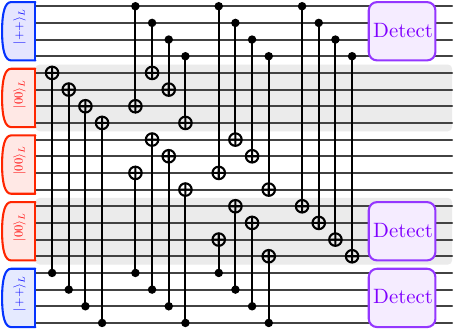}}
\subfigure[]{\includegraphics[scale=.9]{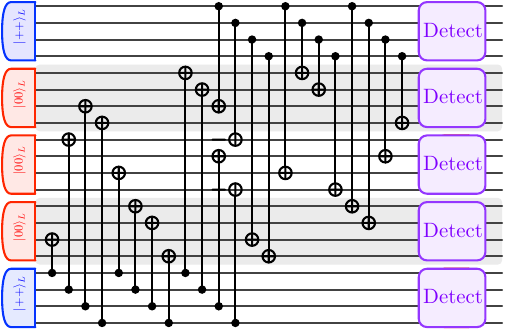}}
\caption{Two circuits for preparing encoded $\ket{00}$ for the $\llbracket 20,2,6 \rrbracket$ code.  Neither circuit is fault tolerant.  The first circuit has a higher acceptance rate and results in comparable logical error and rejection rates for $p \gtrsim 0.002$, but its performance falls off for smaller~$p$ because second-order logical errors begin to dominate.} \label{f:c2026_prep}
\end{figure}

\begin{figure}
\centering
\includegraphics[scale=.5]{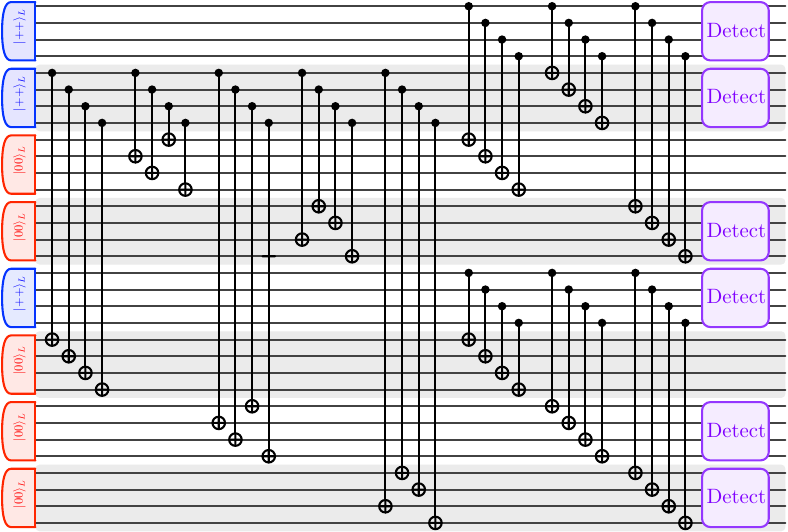}
\caption{Encoding circuit for $\ket{0^4}$ for the $\llbracket 32,4,6 \rrbracket$ code.} \label{f:c3246_prep}
\end{figure}

\begin{figure}
\centering
\subfigure[$\llbracket 36,2,8 \rrbracket \; \ket{00}_L$\label{}]{\raisebox{0cm}{\includegraphics[scale=.4]{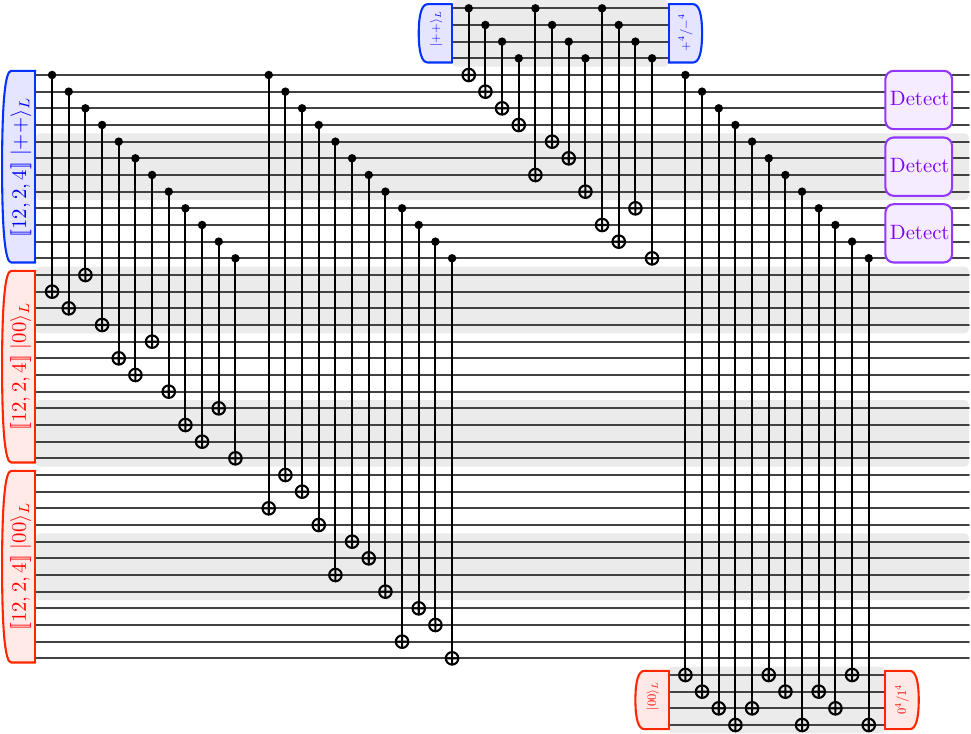}}} \hfill
\subfigure[$\llbracket 48,4,8 \rrbracket \; \ket{0^4}_L$\label{}]{\includegraphics[scale=.4]{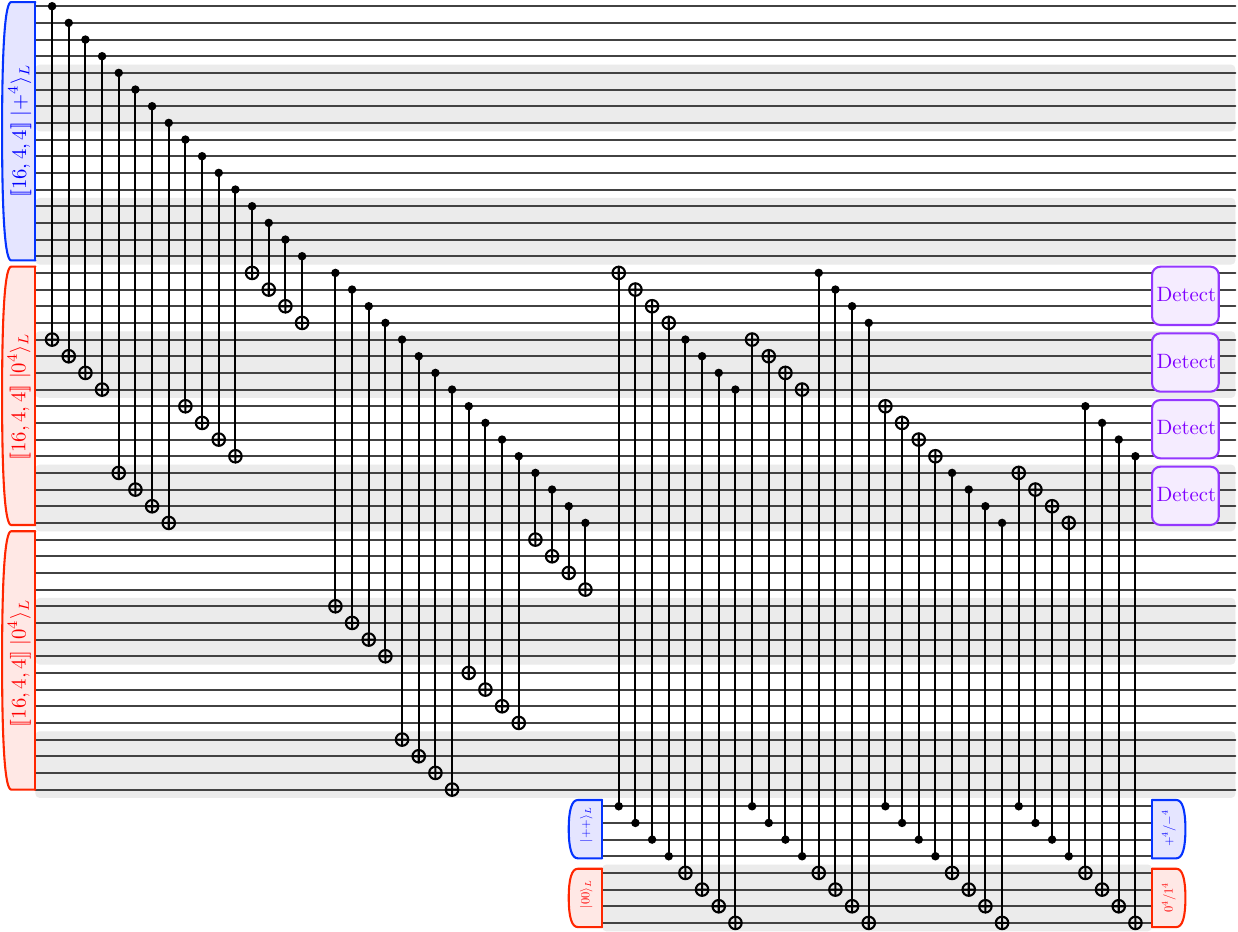}}
\caption{Circuits to prepare and verify encoded $\ket{0^k}$ for the distance-$8$ codes.} \label{f:c3628_c4848_prep}
\end{figure}

\begin{figure}
\centering
\includegraphics[scale=.8]{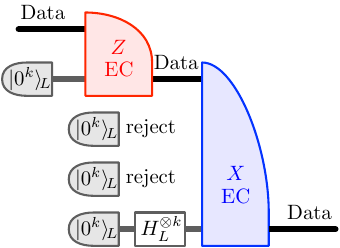}
\caption{An ancilla state factory prepares encoded $\ket{0^k}$ states repeatedly, either sequentially or in parallel.  Accepted states can be used in Steane-style error correction.  In this example, two of the four ancilla states are rejected.} \label{f:statefactory}
\end{figure}

The state-preparation circuits we use for the $\llbracket 20,2,6 \rrbracket$, $\llbracket 32,4,6 \rrbracket$, $\llbracket 36,2,8 \rrbracket$ and $\llbracket 48,4,8 \rrbracket$ codes are given in 
Figs.~\mbox{\ref{f:c2026_prep}--\ref{f:c3628_c4848_prep}}.  
Here the boxed sub-circuits, for the $\llbracket 4,2,2 \rrbracket$, $\llbracket 12,2,4 \rrbracket$ and $\llbracket 16,4,4 \rrbracket$ codes, are given in Figs.~\ref{f:422gadgets} and~\ref{f:12241644prep} in \appref{s:distance5codes}.  
The circuits all have a similar structure.  They begin by preparing encoded $\ket{0^k}$, then they measure certain stabilizers to detect errors.  If any error is detected, then the entire ancilla state is discarded and we must restart the state preparation.  
Acceptance rates are plotted in \figref{f:ancillaacceptanceplot}.  Low ancilla acceptance rates are okay because the ancilla has not yet interacted with the data; only the ancilla preparation needs to be restarted and not the larger quantum circuit.  See \figref{f:statefactory}.  Lower acceptance rates lead to higher overhead, however.  

\begin{figure}
\centering
\subfigure[]{\includegraphics[scale=.8]{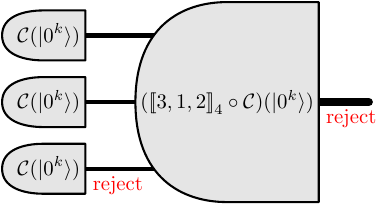}}
\subfigure[]{\includegraphics[scale=.8]{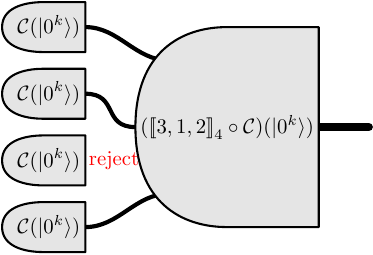}}
\caption{The circuit to prepare $\ket{0^k}$ encoded in the $\llbracket 3,1,2 \rrbracket_4 \circ {\cal C}$ code takes as input three $\ket{0^k}$ states encoded in ${\cal C}$; see \figref{f:c3628_c4848_prep}.  (a) A one-stage ancilla factory works non-adaptively.  If an input ${\cal C}$-encoded state is rejected, then so is the full output state, meaning that a substantial amount of work is wasted.  (b) In a two-stage factory, the $\llbracket 3,1,2 \rrbracket_4 \circ {\cal C}$ state preparation selects or waits for three accepted ${\cal C}$ states.  This requires adaptive control, but reduces the overhead, as shown in \figref{f:ancillaacceptance}.} \label{f:factorytwostage}
\end{figure}

The $\llbracket 36,2,8 \rrbracket$ and $\llbracket 48,4,8 \rrbracket$ codes are $\llbracket 3,1,2 \rrbracket_4 \circ \cC$, for $\cC$ either $\llbracket 12,2,4 \rrbracket$ or $\llbracket 16,4,4 \rrbracket$.  Each state-preparation circuit begins by preparing three copies of encoded $\ket{0^k}$ or $\ket{+^k}$ in $\cC$.  We can therefore consider a two-stage ancilla factory, where the first stage outputs accepted $\cC$-encoded states and the second stage adaptively takes three of these and combines them into a $\llbracket 3,1,2 \rrbracket_4 \circ \cC$-encoded state.  See \figref{f:factorytwostage}.  This adaptive control reduces the wasted work when a $\cC$-encoded state is rejected, therefore reducing the overhead.  The dashed lines in \figref{f:ancillaacceptanceplot} give the acceptance probability of the second stage conditioned on being given three accepted $\cC$-encoded states.  The dashed lines in \figref{f:ecnotsplot} give the expected number of CNOT gates needed to prepare $\llbracket 3,1,2 \rrbracket_4 \circ \cC$-encoded $\ket{0^k}$ states assuming two-stage adaptive control.  At $p = 0.4\%$, using a two-stage factory reduces the expected number of CNOTs by $30.3\%$ and $37.6\%$ for the $\llbracket 36,2,8 \rrbracket$ and $\llbracket 48,4,8 \rrbracket$ codes, respectively, with smaller savings for smaller~$p$.  

Consider the two circuits for preparing $\llbracket 20,2,6 \rrbracket$-encoded $\ket{0^2}$ in \figref{f:c2026_prep}.  Neither circuit is fault tolerant.  In fact, in both circuits, a second-order fault (e.g., $X$ on the third and fourth qubits) can lead to an undetected weight-four error that decodes to a logical error.  However, circuit (a) has several times as many bad second-order faults.  At high noise rates~$p$, this does not matter much, but for small~$p$ the second-order faults begin to dominate and circuit (b) performs significantly better.  This is a tradeoff, however, since the smaller circuit (a) has lower overhead, e.g., $24.0\%$ fewer CNOT gates in expectation at $p = 0.4\%$.  

\begin{figure}
\newcommand{\inc}[1]{\includegraphics[scale=.7]{images/histograms/#1}}
\begin{gather*}
\begin{array}{l | c c c c}
\hline \hline
                   & \multicolumn{4}{c}{p} \\
\text{Code} & .0005 & .001 & .002 & .004 \\ \hline
\llbracket 20,2,6 \rrbracket \; (\text{a}) & \inc{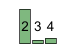} & \inc{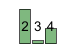} & \inc{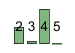} & \inc{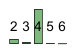} \\
\llbracket 20,2,6 \rrbracket \; (\text{b}) & \inc{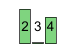} & \inc{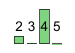} & \inc{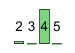} & \inc{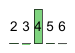} \\
\llbracket 32,4,6 \rrbracket & \inc{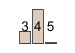} & \inc{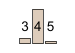} & \inc{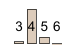} & \inc{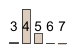} \\[.15cm]
\llbracket 36,2,8 \rrbracket & \inc{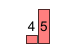} & \inc{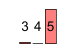} & \inc{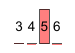} & \inc{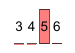} \\
\llbracket 48,4,8 \rrbracket & \inc{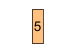} & \inc{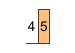} & \inc{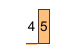} & \inc{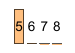} \\
\hline \hline
\end{array} \\
\text{(a) Logical error orders} \\[.2cm]
\begin{array}{l | c c c c}
\hline \hline
                   & \multicolumn{4}{c}{p} \\
\text{Code} & .0005 & .001 & .002 & .004 \\ \hline
\llbracket 20,2,6 \rrbracket \; (\text{a}) & \inc{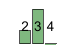} & \inc{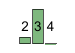} & \inc{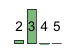} & \inc{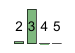} \\
\llbracket 20,2,6 \rrbracket \; (\text{b}) & \inc{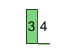} & \inc{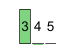} & \inc{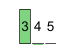} & \inc{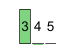} \\
\llbracket 32,4,6 \rrbracket & \inc{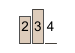} & \inc{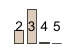} & \inc{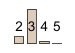} & \inc{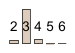} \\[.15cm]
\llbracket 36,2,8 \rrbracket & \inc{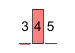} & \inc{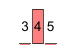} & \inc{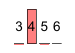} & \inc{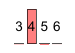} \\
\llbracket 48,4,8 \rrbracket & \inc{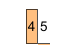} & \inc{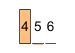} & \inc{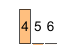} & \inc{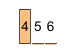} \\
\hline \hline
\end{array} \\
\text{(b) Rejection orders}
\end{gather*}
\caption{Histograms of the number of faults causing each logical error or rejection, at different noise rates~$p$.  Were the distance-$d$ code circuits fault tolerant, a logical error would require $d/2+1$ or more faults, and a rejection would require at least $d/2$ faults.  Observe that none of the circuits are fault tolerant.  At high~$p$, the respective histograms are dominated by $d/2+1$ and $d/2$ faults.  For smaller~$p$, lower-order faults become relatively more likely.  
} \label{f:histograms}
\end{figure}

In order to judge the effective fault tolerance of the preparation circuits, \figref{f:histograms} gives histograms of the number of faults causing each logical error or rejection, for different values of~$p$.  To obtain this data, having sampled a set of faults causing a logical error or rejection, we find a minimal subset of faults causing the same result and record the size of that minimal set.  Although none of the circuits are fault tolerant to distance~$d$, the logical error rates and rejection rates do appear to go as $p^{d/2+1}$ and $p^{d/2}$, respectively, for large values of~$p$, because the higher-order faults combinatorially dominate.  For smaller~$p$, lower-order faults become relatively more likely.  
For example, the large fraction of order-two logical errors at $p = 5 \times 10^{-4}$ for the $\llbracket 20,2,6 \rrbracket$ code circuit (a) explains why the corresponding curve in \figref{f:logicalerrorplot} flattens from cubic to quadratic $p$ dependence for small~$p$.  The histogram suggests that the logical error rate for the $\llbracket 20,2,6 \rrbracket$ code circuit (b) will begin to flatten for $p < 5 \times 10^{-4}$.  

These examples demonstrate that circuits that are not strictly fault tolerant can still perform very well, particularly at higher noise rates.  

Where do the circuits in Figs.~\ref{f:c2026_prep}--\ref{f:c3628_c4848_prep} come from?  Each circuit has two parts, beginning with a subcircuit encoding $\ket{0^k}$ and ending with a verification subcircuit to catch bad errors.  The error-detection subcircuits have been constructed heuristically, according to intuition and simulations.  The encoding subcircuits generalize a method of Steane~\cite{Steane02}.  In Steane's method, Gaussian elimination is applied to the matrix of $X$ stabilizers of a CSS code, then CNOT gates are applied from the ``pivot" qubits prepared in $\ket{+}$ to the ``free" qubits prepared in $\ket{0}$.  As an example, the $\llbracket 7,1,3 \rrbracket$ self-dual CSS code has the following $X$ stabilizers and $\ket{0}$ encoding circuit: 
\begin{equation*}
\def\I{\cdot}
\definecolor{myRed}{rgb}{1,0,0}
\definecolor{myGreen}{rgb}{0,.8,0}
\definecolor{myBlue}{rgb}{0,0,1}
\def\rX{{\color{myRed}X}}
\def\gX{{\color{myGreen}X}}
\def\bX{{\color{myBlue}X}}
\begin{array}{r @{\;} c @{\;} c @{\;} c @{\;} c @{\;} c @{\;} c @{\;} c}
&\I&\I&\I&\rX&\rX&\rX&\rX\\
&\I&\gX&\gX&\I&\I&\gX&\gX\\
&\bX&\I&\bX&\I&\bX&\I&\bX
\end{array}
\qquad\quad
\raisebox{-.9cm}{\includegraphics[scale=1]{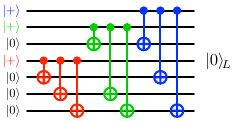}}
\end{equation*}
(The CNOTs' order does not matter in the noiseless case.)  

We generalize the above method to apply to codes concatenated on the $\llbracket 4,2,2 \rrbracket$ code.  It is more complicated because the $\llbracket 4,2,2 \rrbracket$ code encodes two logical qubits, and transversal CNOT gates between two code blocks implements two logical CNOT gates.  
Recall $\D_X$ mapping $X \mapsto XI$, $Z \mapsto IX$.  Then much as the CNOT gate copies an $X$ stabilizer forward, 
\begin{equation*}
\includegraphics[scale=1]{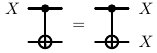}
\end{equation*}
transversal CNOTs between two $\llbracket 4,2,2 \rrbracket$ code blocks copy both $\D_X(X)_L = XXII$ and $\D_X(Z)_L = XIXI$ forward: 
\begin{gather*}
\includegraphics[scale=.667]{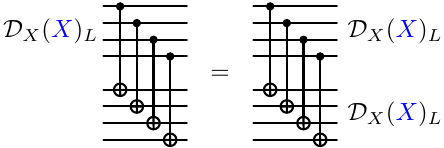} \\
\includegraphics[scale=.667]{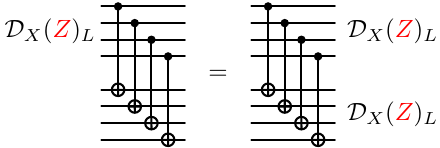}
\end{gather*}
By permuting the qubits in the target code block, we can copy $\D_X(X)$ and $\D_X(Z)$ into other logical operators.  For example, 
\begin{gather*}
\includegraphics[scale=.667]{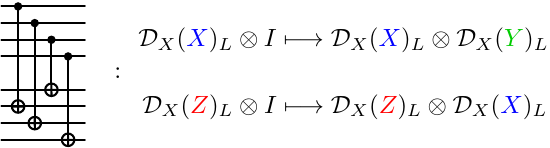} \\
\includegraphics[scale=.667]{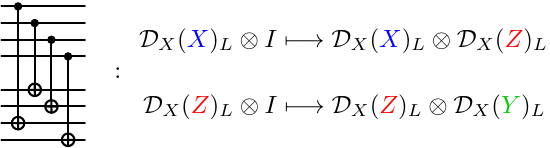}
\end{gather*}
Similar circuits allow us to copy $(X,Z)$ into any of $\{ (X,Y), (X,Z), (Y,X), (Y,Z), (Z,X), (Z,Y) \}$.  
These circuits have been used repeatedly in Figs.~\ref{f:c2026_prep} and~\ref{f:c3246_prep}, e.g., in \figref{f:c2026_prep} the transversal CNOT gates copy $\D_X(XIIII)_L$ and $\D_X(ZIIII)_L$ into $\D_X(XZZXI)_L$ and $\D_X(ZYYZI)_L$, respectively, and copy $\D_X(IIIIX)_L$ and $\D_X(IIIIZ)_L$ into $\D_X(IXZZX)_L$ and $\D_X(IZYYZ)_L$, respectively.  

The above transversal CNOT circuits, building out two $X$ stabilizers at a time, motivates the following definition: 

\begin{definition} \label{t:pairedsupport}
Pauli operators $P$ and $Q$ have \emph{paired support} if $P$, $Q$ and $P \cdot Q$ all have the same support.  
A set of Pauli operators has paired support if it can be partitioned into pairs that each have paired support.  
\end{definition}

The condition that $P \cdot Q$ has the same support as $P$ and $Q$ means that $P$ and $Q$ have different Pauli operators on each qubit in their joint support.  This means that $\D_X(P)_L$ and $\D_X(Q)_L$ can be built out simultaneously using transversal CNOT gates between $\llbracket 4,2,2 \rrbracket$ code blocks.  The paired support property is useful because building two $X$ stabilizers together can be more efficient than building them up separately as in Steane's original method.  (Ref.~\cite{Paetznickgolay2013} has previously considered sharing CNOT gates between $X$ stabilizers in order to simplify encoding circuits.)

As examples of \defref{t:pairedsupport}, the $\llbracket 5,1,3 \rrbracket$ and $\llbracket 8,2,3 \rrbracket$ codes in Figs.~\ref{f:2026code}(a) and~\ref{f:823code} are Hermitian, or GF(4)-linear, i.e., invariant when the Paulis are permuted by $(XYZ)$~\cite{CrossVandeth25smallcodes}.  Every GF(4)-linear stabilizer code has stabilizer generators with paired support.  The converse statement is not true, however.  

Other distance-three GF(4)-linear codes that might be of interest, especially for quantum memory, include $\llbracket 9,3,3 \rrbracket$, $\llbracket 10,4,3 \rrbracket$, $\llbracket 11,5,3 \rrbracket$, $\ldots$, $\llbracket 16,10,3 \rrbracket$~\cite{Grassl07codetable}.  We have only considered $\llbracket 5,1,3 \rrbracket \circ \llbracket 4,2,2 \rrbracket$ and $\llbracket 8,2,3 \rrbracket \circ \llbracket 4,2,2 \rrbracket$ because computation on the encoded data becomes more difficult with more logical qubits per code block.  No $\llbracket 8,3,3 \rrbracket$ code has a paired support presentation of its stabilizer generators.  

As to distance-four codes, there is a $\llbracket 10,2,4 \rrbracket$ code with paired support, that is not GF(4)-linear, and there are GF(4)-linear codes with parameters $\llbracket 12,4,4 \rrbracket$, $\llbracket 14,6,4 \rrbracket$, $\llbracket 17,9,4 \rrbracket$, $\llbracket 20,10,4 \rrbracket$, and so on~\cite{Grassl07codetable}.  Concatenating with $\llbracket 4,2,2 \rrbracket$, we obtain $\llbracket 40,4,8 \rrbracket$ and $\llbracket 48,8,8 \rrbracket$ codes, for instance, that might be competitive with the $\llbracket 3,1,2 \rrbracket_4^{\circ 2} \circ \llbracket 4,2,2 \rrbracket = \llbracket 48,4,8 \rrbracket$ code that we simulated above.  We have focused on the $\llbracket 48,4,8 \rrbracket$ code because of its simple concatenated structure.  

The smallest distance-five code, $\llbracket 11,1,5 \rrbracket$, has paired support, even though it is not GF(4)-linear~\cite{Gottesman97thesis}.  Other potentially interesting distance-five codes include $\llbracket 13,1,5 \rrbracket$ and $\llbracket 15,3,5 \rrbracket$ cyclic, GF(4)-linear codes~\cite{VatanRoychowdhuryAnantram97codes, GrasslBeth99bchcodes}.  We have not simulated these codes concatenated on $\llbracket 4,2,2 \rrbracket$, because they have lower rates, distance-$10$ fault tolerance is more difficult to achieve, and because distance-eight protection is likely sufficient for near-term quantum computers.

\section{Logical Clifford gates}

So far we have studied error correction, which is needed for maintaining a protected quantum memory.  Our goal is protected quantum \emph{computation}.  We achieve this by supplementing logical Hadamard and CNOT gates, real Clifford operations, with non-Clifford $Z$ rotations.  In this section, we explain the logical Clifford operations.  We reduce the analysis to transversal logical CNOT gates between two code blocks.  

As the codes we consider are CSS, transversal CNOT gates between two code blocks implement logical transversal CNOT gates.  

For the $\llbracket 4,2,2 \rrbracket$, $\llbracket 12,2,4 \rrbracket$, $\llbracket 20,2,6 \rrbracket$ and $\llbracket 36,2,8 \rrbracket$ codes, simply permuting the qubits in a code block can implement logical CNOT gates in both directions.  
Together with two rounds of transversal CNOT gates, this allows for targeted logical CNOT gates between two code blocks.  At the logical level, we have the identity: 
\begin{equation*}
\includegraphics[scale=1]{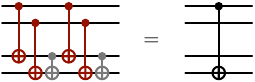}
\end{equation*}
Here the red CNOT gates are transversal, and the gray logical sub-circuits are generated by qubit permutations.  Similar identities allow for targeting either of the bottom two logical qubits with a logical CNOT controlled by either of the top two logical qubits.  Note that error correction should be applied between the rounds of transversal CNOT gates, in order to prevent weight-one errors in one block from spreading to weight-two errors in the other.  

For the $\llbracket 16,4,4 \rrbracket$, $\llbracket 32,4,6 \rrbracket$ and $\llbracket 48,4,8 \rrbracket$ codes, qubit permutations of a code block can implement any of the $36$ logical operations generated by 
\begin{equation*}
\includegraphics[scale=1]{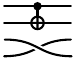} \qquad
\includegraphics[scale=1]{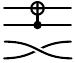} \qquad
\includegraphics[scale=1]{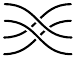}
\end{equation*}
Together with four rounds of transversal CNOT gates, this allows for targeted logical CNOT gates between two code blocks, via the logical-level identity: 
\begin{equation*}
\includegraphics[scale=.9]{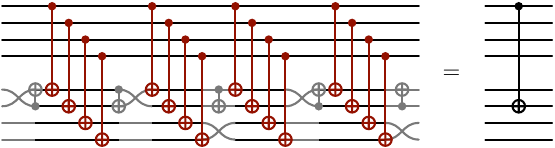}
\end{equation*}
Similar identities allow for targeting any of the bottom four logical qubits with a logical CNOT controlled by any of the top four logical qubits.  Again, error correction should be applied between the rounds of transversal CNOT gates.  

It is curious that for the $k$ logical qubit codes, $k \in \{2, 4\}$, applying $k$ logical CNOTs between code blocks is $k$ times \emph{easier} than applying one logical CNOT.  The exact opposite holds for physical qubits.  

Up to qubit permutations, transversal Hadamard gates on a code block implements transversal logical Hadamard gates.  With good one-qubit gates, the logical error and rejection rates for logical Hadamards followed by error correction will not be significantly higher than for error correction alone.  Targeted logical Hadamard gates can be implemented, for example, by moving the target logical qubit into a different code block and applying Hadamards there.  Targeted logical CNOTs and Hadamards generate the real Clifford group.  

Since in our model, qubit permutations and Hadamards are free, the performance of noisy real Clifford operations comes down to the performance of transversal logical CNOT gates.  We study two ways of implementing them.  

\begin{figure}
\centering
\subfigure[\label{f:transversalcnotsimulation}]{\includegraphics[scale=1]{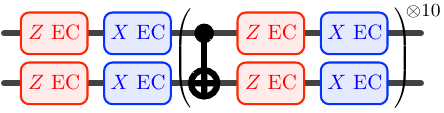}}
\subfigure[\label{f:teleportedcnotsimulation}]{\includegraphics[scale=.84]{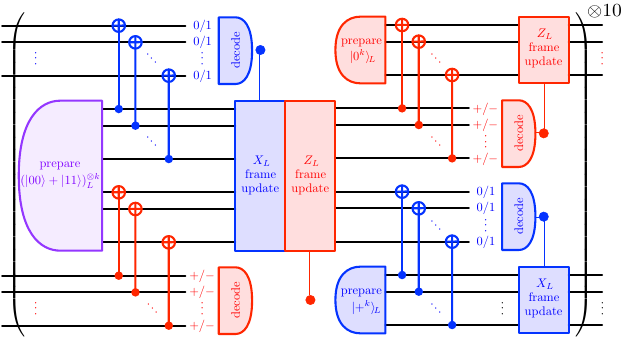}}
\caption{Two circuits we simulate in order to estimate the logical error and rejection rates for transversal logical CNOT gates.  The first circuit only requires prepared $\ket{0^k}_L$ and $\ket{+^k}_L$ ancilla states in one code block.  The second circuit requires a two-code-block ancila state $(\ket{00}+\ket{11})_L^{\otimes k}$, which has a higher preparation overhead.} \label{f:transversalteleportedcnotsimulation}
\end{figure}

In our first circuit to estimate the logical error and rejection rates for transversal CNOT gates, shown in \figref{f:transversalcnotsimulation}, we simulate ten rounds of transversal CNOT gates between two code blocks, sandwiched between error correction operations.  
In the simulation, we assume that the inputs are perfect, and finish with perfect decoding.  We divide the total logical error rate (including all combinations of $X$, $Y$ and $Z$ errors) and rejection rate by $10$ to get per-round rates.  

Figure~\ref{f:teleportedcnotsimulation} shows our second circuit for implementing and simulating transversal logical CNOT gates.  This circuit is based on the following identity for teleporting into a CNOT gate using a prepared cat state $\ket{00} + \ket{11}$: 
\begin{equation*}
\includegraphics[scale=1]{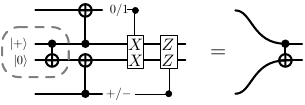}
\end{equation*}
When run at the logical level, using a prepared logical $(\ket{00} + \ket{11})^{\otimes k}$ state, the control block is automatically $X$ error corrected and the target block automatically $Z$ error corrected.  Therefore we insert $Z$ error correction on the control block and $X$ error correction on the target block.  We repeat everything ten times, and divide the total logical error rate and rejection rate by $10$.  

In order to prepare $(\ket{00} + \ket{11})_L^{\otimes k}$ reliably, we can prepare one code block with $\ket{+^k}_L$, another with $\ket{0^k}_L$, and apply transversal CNOT gates between them.  We use the circuits from Figs.~\ref{f:c2026_prep} to~\ref{f:c3628_c4848_prep}, except delay the final error detections until after the transversal CNOTs.  

\begin{figure}
\centering
\subfigure[$\llbracket 20,2,6 \rrbracket$ code \label{f:c2026cnots}]{\includegraphics[scale=.41]{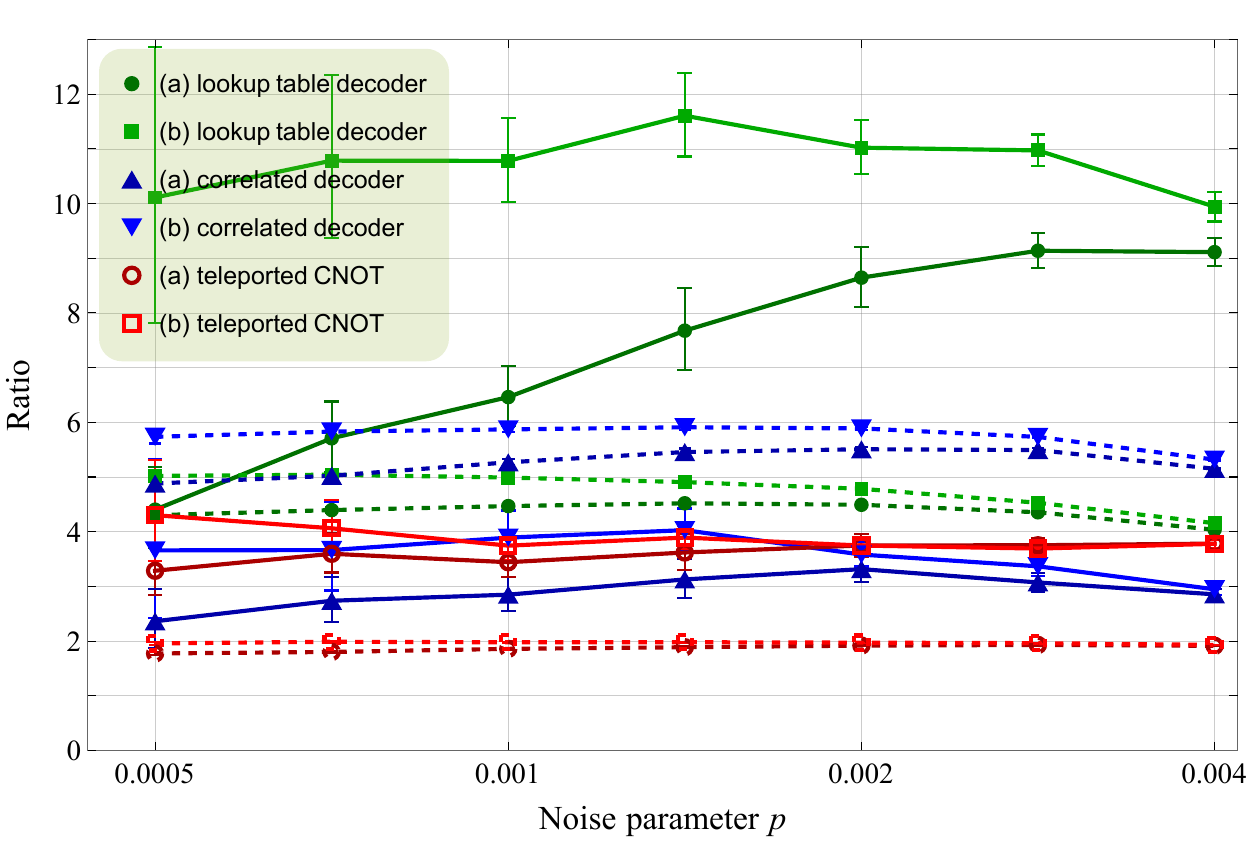}}
\subfigure[$\llbracket 36,2,8 \rrbracket$ code \label{f:c3628cnots}]{\includegraphics[scale=.41]{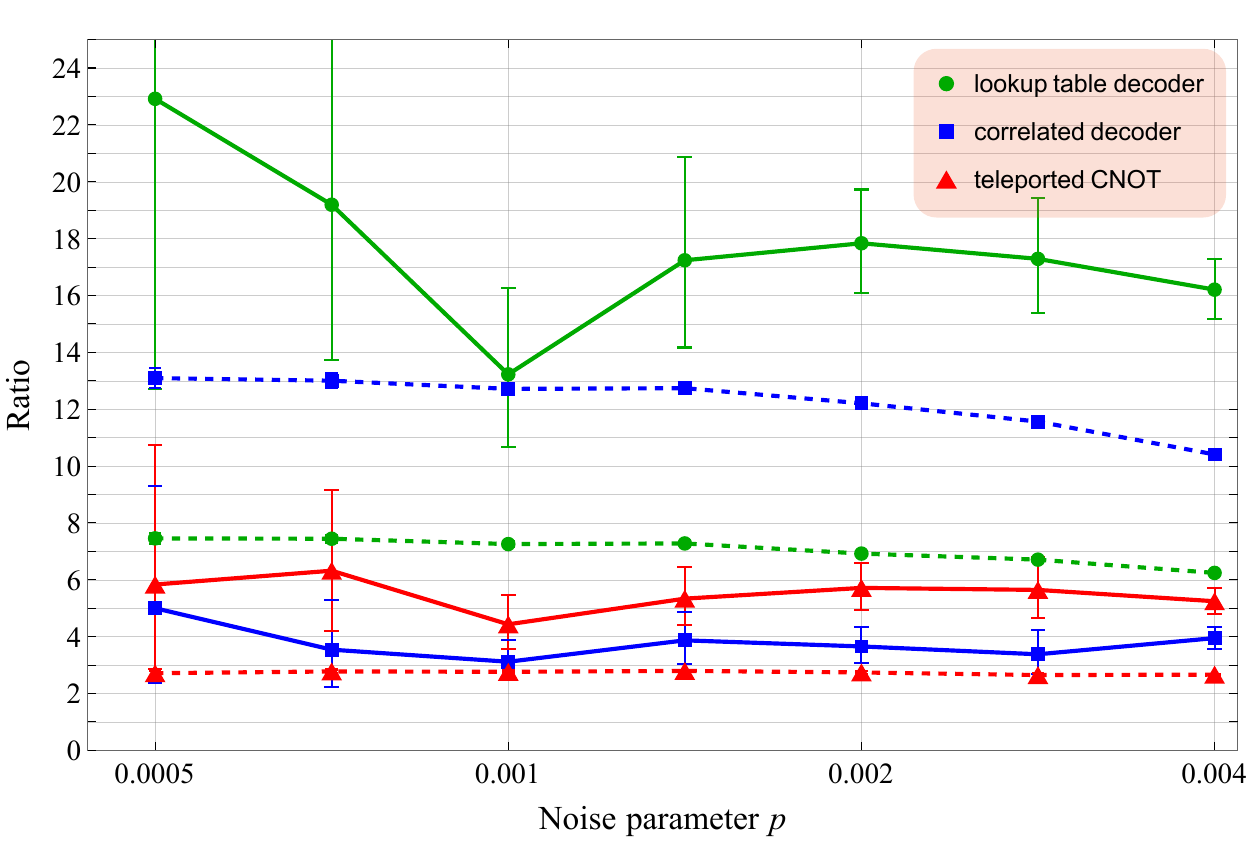}}
\caption{Transversal CNOT gates, with error correction, have higher logical error and rejection rates than error correction alone.  
Here for the $\llbracket 20,2,6 \rrbracket$ and $\llbracket 36,2,8 \rrbracket$ codes, we plot the \emph{ratios} between the logical error rate in our transversal CNOT simulation and that in our repeated error correction simulation (solid lines), and the similar ratios between rejection rates (dashed lines).  
For the $\llbracket 20,2,6 \rrbracket$ code, we consider the two circuits from \figref{f:c2026_prep}; the second circuit is more reliable but has higher overhead.  
Note that the repeated error correction simulation involved only one code block, not two, and only measured only logical $X$ error rates, not $X$ and $Z$.  
} \label{f:cnotsratios}
\end{figure}

We focus our simulations on the promising $\llbracket 20,2,6 \rrbracket$ and $\llbracket 36,2,8 \rrbracket$ codes.  Figure~\ref{f:cnotsratios} shows the ratios between logical error rates (solid lines) and rejection rates (dashed lines) for the CNOT and repeated error correction simulations.  

Consider the green curves in \figref{f:cnotsratios}.  For these curves, the transversal CNOT simulation follows \figref{f:transversalcnotsimulation}, where the error correction procedures are exactly the same---a simple lookup table---as used in the repeated error correction simulation.  The performance is not impressive, with logical error rates especially high.  Intuitively, since the simulation involves two code blocks, the rejection rate per round of CNOT gates should be at least $2$ to $2.2$ times the per-round rejection rate for repeated error correction ($2.2$ because there are $11$ rounds of error correction and we divide by $10$).  The logical error rate per round of CNOT gates should be at least $4$ to $4.4$ times the per-round logical error rate for repeated error correction.  Here one factor of two is from the two code blocks, and another factor of two because our repeated error correction simulation only measures logical $X$ errors, not~$Z$ errors.  
In fact, though, rejection and logical error rates are much worse.  The CNOT gates copy $X$ errors downward, for example, so immediately after the CNOT gates, the target block has physical $X$ error rate roughly twice what it would be without the CNOTs.  As rejection rates scale roughly as $c' p^{d/2}$ and logical error rates as $c \, p^{d/2+1}$, doubling $p$ can increase these rates by $2^{d/2}$ and $2^{d/2+1}$, respectively.  This explains the very high solid green error rate ratios.  (The error ratio for the $\llbracket 20,2,6 \rrbracket$ circuit (a) is lower at small~$p$ because low-order logical errors dominate, as in \figref{f:histograms}.)  

The blue curves in \figref{f:cnotsratios} are also for the transversal CNOT simulation from \figref{f:transversalcnotsimulation}, except with a heuristic ``correlated" decoder that rejects the most common fault patterns between nearby $X$ or $Z$ error correction modules.  This reduces the logical error rate substantially, at the cost of increasing the rejection rate.  It works particularly well for the $\llbracket 20,2,6 \rrbracket$ code, for which the logical error rate ratio drops well below $4$ for large~$p$.  This is possible because the CNOTs copy errors between code blocks, so a joint decoding of the code blocks can take advantage of more information about the likely errors.  

\begin{figure}
\centering
\includegraphics[scale=.41]{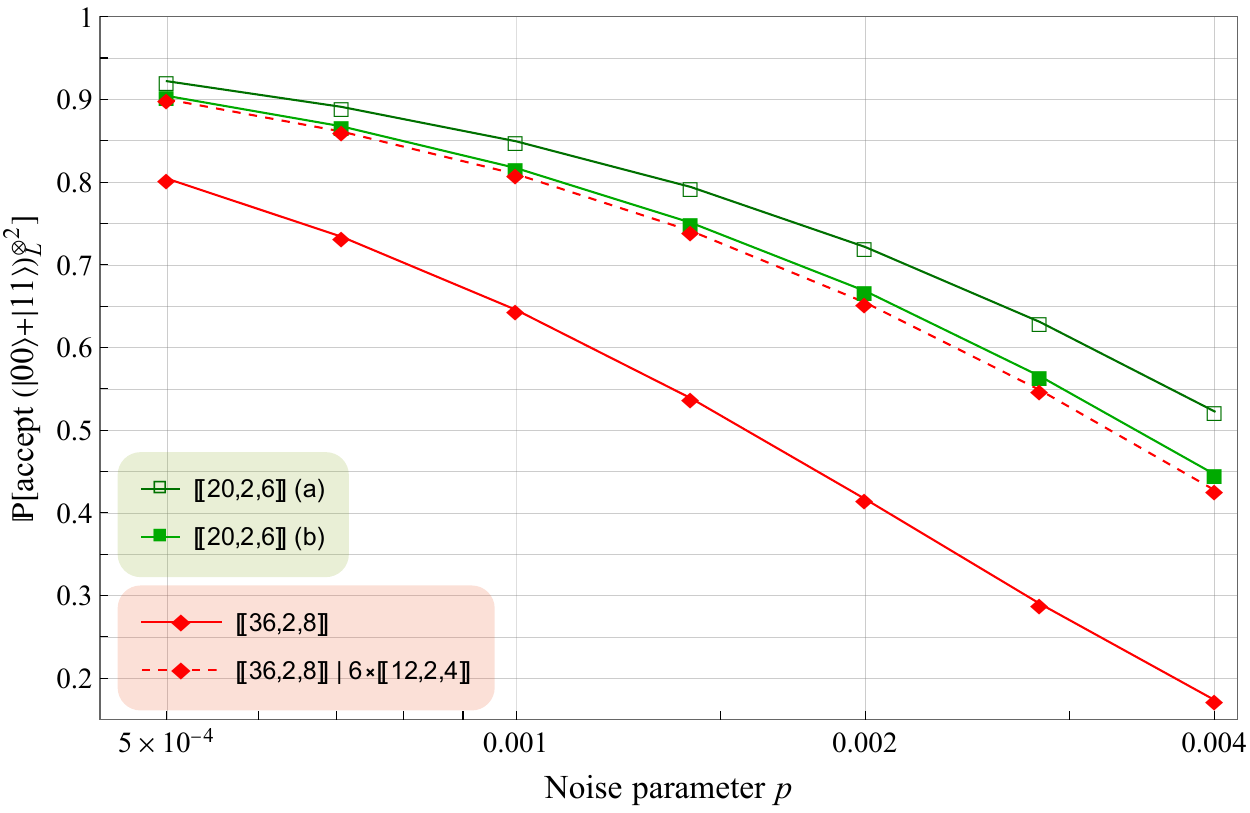}
\caption{Acceptance probabilities for preparing $(\ket{00}+\ket{11})_L^{\otimes 2}$.  The dashed line is for a two-stage ancilla factory; it gives the acceptance rates for the $\llbracket 36,2,8 \rrbracket$ code conditioned on being given six accepted copies of $\ket{0^k}_L$ for the $\llbracket 12,2,4 \rrbracket$ code.  
} \label{f:bellancillaacceptanceplot}
\end{figure}

Overall, across the range of noise parameters, teleporting into logical CNOT gates gives the lowest rejection rate, and almost the lowest logical error rate (red curves), even using the naive lookup table decoder.  However, this method has a higher state preparation overhead, since preparing a verified logical $(\ket{00} + \ket{11})^{\otimes k}$ state has higher overhead than preparing two copies of $\ket{0^k}_L$.  Figure~\ref{f:bellancillaacceptanceplot} plots ancilla acceptance rates for one- and two-stage ancilla factories.

\section{Non-Clifford operations}

For small angles, reliable logical $Z$ rotations can be implemented using physical $Z$ rotations and error detection~\cite{ChoiChongEnglundDing23zrotations}.  Arbitrary-angle rotations can be implemented by preparing logical Bell pairs between the $\llbracket 4,2,2 \rrbracket$ and, e.g., $\llbracket 20,2,6 \rrbracket$ codes, using physical two-qubit gates to implement the logical rotation on the $\llbracket 4,2,2 \rrbracket$ half~\cite{HeAmaroShaydulinPistoia24iceberg}, then teleporting into the other code.  If necessary, non-stabilizer ``magic" states can be distilled to get higher reliability~\cite{BravyiKitaev04magic, GuptaIBM24magiccz}, but this will incur overhead from distillation and rotation gate synthesis.  We leave to future work the analysis of error rates and overhead for these different methods of achieving quantum universality.

\section*{Acknowledgments}

We thank Marcus Silva, Zulfi Alam and Matthias Troyer for their support and feedback during this project.  
R.C.\ is currently at NVIDIA; all research was conducted while at Microsoft.  

\appendix

\section{Codes with distance $\leq 5$} \label{s:distance5codes}

\begin{figure}
\centering
\subfigure[\label{f:lowdistancecodeslogicalerror}]{
\includegraphics[scale=.48]{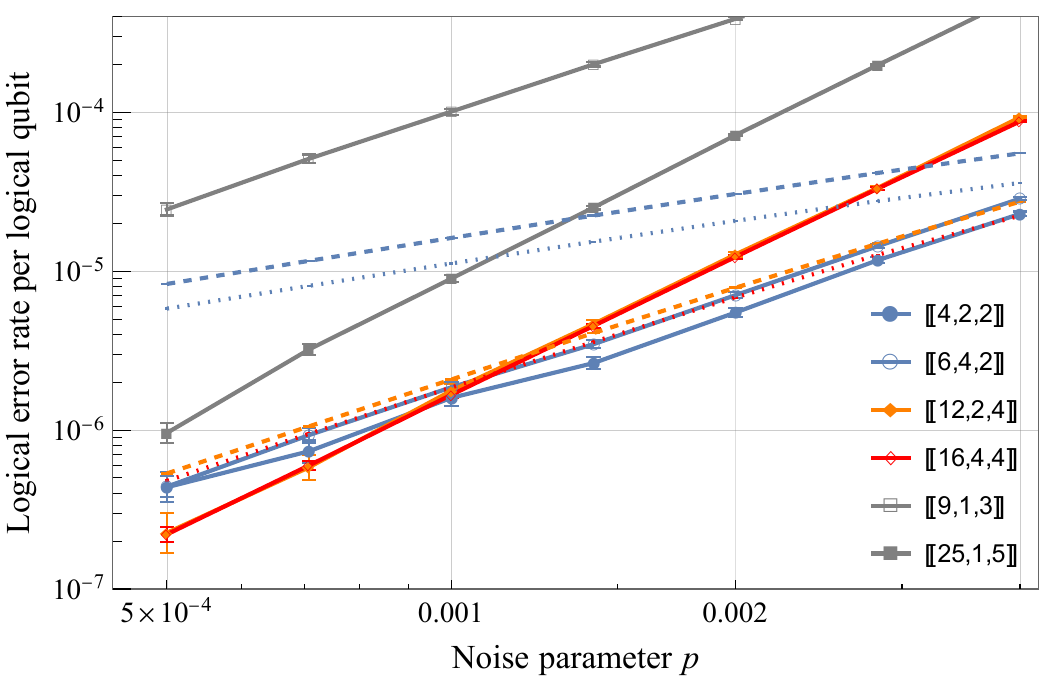}
}
\subfigure[]{
\includegraphics[scale=.48]{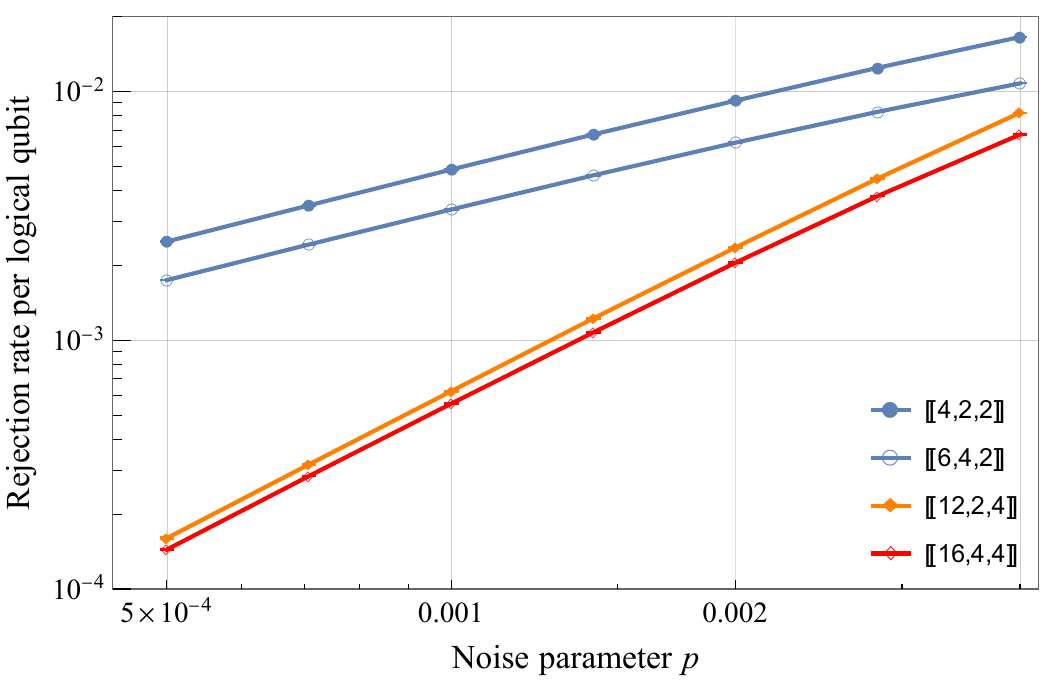}
}
\caption{Logical error rates and rejection rates per logical qubit over one round of Steane-style error correction for the $\llbracket 12,2,4 \rrbracket$ and $\llbracket 16,4,4 \rrbracket$ codes (orange and red), one round of Shor-style error-correction for the $\llbracket 4,2,2 \rrbracket$ and $\llbracket 6,4,2 \rrbracket$ Iceberg codes, and one cycle of syndrome extraction for $\llbracket d^2, 1, d \rrbracket$ surface codes, $d = 3, 5$.  Note that fault-tolerant error correction for the surface codes takes $d$ cycles of syndrome extraction.} \label{f:lowdistancecodes}
\end{figure}

\begin{figure}[b]
\centering
\includegraphics[scale=1]{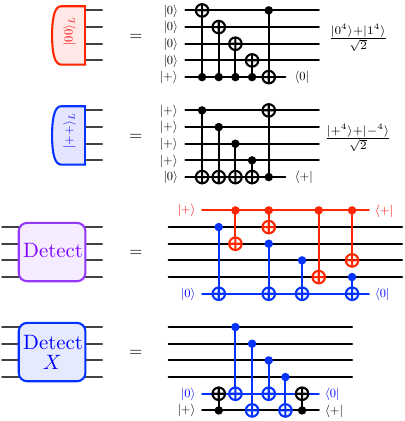}
\caption{Fault-tolerant state preparation and error-detection gadgets for the $\llbracket 4,2,2 \rrbracket$ code.} \label{f:422gadgets}
\end{figure}

\begin{figure}
\centering
\subfigure[$\llbracket 12,2,4 \rrbracket \; \ket{0^2}_L$\label{f:1224prep}]{\includegraphics[scale=1]{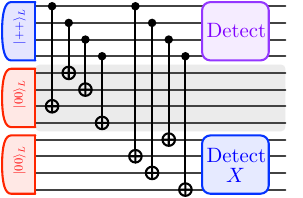}}
\subfigure[$\llbracket 16,4,4 \rrbracket \; \ket{0^4}_L$\label{f:1644prep}]{\includegraphics[scale=.9]{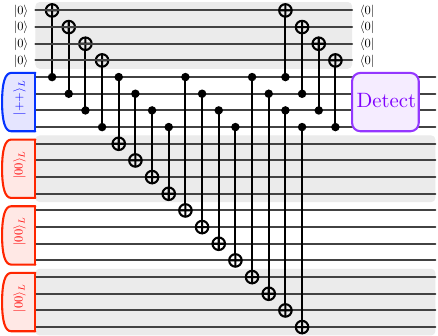}}
\caption{Fault-tolerant circuits to prepare encoded $\ket{0^k}$ for the $\llbracket 12,2,4 \rrbracket$~\cite{Silva24microsoft12qubitcode} and $\llbracket 16,4,4 \rrbracket$ codes.  In each, the state is accepted if no error is detected.  Figure~\ref{f:ancillaacceptance} plots acceptance rates.  
} \label{f:12241644prep}
\end{figure}

Figure~\ref{f:memorynoiseplot}, in the code capacity model, suggests that at moderate noise rates, low-distance codes will have trouble achieving logical error rates below $\sim 10^{-6}$.  In this appendix, we substantiate this conclusion using circuit-noise simulations.  

Figure~\ref{f:lowdistancecodes} shows the results of simulating the $\llbracket 4,2,2 \rrbracket$ and $\llbracket 6,4,2 \rrbracket$ Iceberg codes, $\llbracket 3,1,2 \rrbracket_4 \circ \llbracket 4,2,2 \rrbracket = \llbracket 12,2,4 \rrbracket$ Knill code, $\llbracket 4,2,2 \rrbracket^{\circ 2} = \llbracket 16,4,4 \rrbracket$ code, and $\llbracket d^2, 1, d \rrbracket$ surface codes for $d \in \{3, 5\}$.  
All simulations use the stochastic noise model from \defref{t:noise}.  
Other than for the surface codes, we simulate state preparation, $10$ rounds of error detection or correction, and transversal measurement.  
For the $\llbracket n,n-2,2 \rrbracket$ Iceberg codes, we do not use Steane-style error correction, but rather much more qubit- and gate-efficient fault-tolerant error-detection circuits from~\cite{ChaoReichardt17errorcorrection, self22icebergcode}; see \figref{f:422gadgets}.  
For the $\llbracket 12,2,4 \rrbracket$ and $\llbracket 16,4,4 \rrbracket$ codes, we use Steane-style error correction, with the fault-tolerant state preparation circuits from \figref{f:12241644prep}.  
For the surface codes, we simulate $10$ cycles of noisy syndrome extraction.  

Observe that the surface codes have unfavorable logical error rates compared to the other codes; however they also have $0$ rejection rate.  
For moderate $p$, the Iceberg codes have lower logical error rates than the higher-distance codes.  Unfortunately, they have very high rejection rates, e.g., for the $\llbracket 4,2,2 \rrbracket$ code at $p = 0.004$, $p_R / p_L = 722^{+26}_{-25}$, and it is greater for smaller~$p$.  
The dashed and dotted blue lines in \figref{f:lowdistancecodeslogicalerror} plot $p_R / (300 k)$ for the $\llbracket 4,2,2 \rrbracket$ and $\llbracket 6,4,2 \rrbracket$ codes, respectively.  
Running error detection less frequently should help this unfavorable ratio.  It might also be acceptable for applications requiring low total logical error rates.  

The normalized logical error and rejection rates, $p_L / k$ and $p_R / k$, for the $\llbracket 12,2,4 \rrbracket$ and $\llbracket 16,4,4 \rrbracket$ codes are difficult to distinguish.  
For both codes, $p_R / p_L$ breaks $300$ at $p = 0.001$, at which for $\llbracket 12,2,4 \rrbracket$, $p_R / p_L = 353^{+39}_{-35}$ and for $\llbracket 16,4,4 \rrbracket$, $p_R / p_L = 334(13)$.  
The dashed orange and dotted red lines in \figref{f:lowdistancecodeslogicalerror} plot $p_R / (300 k)$ for the $\llbracket 12,2,4 \rrbracket$ and $\llbracket 16,4,4 \rrbracket$ codes, respectively.  

\clearpage
\bibliographystyle{halpha-abbrv}
\bibliography{bib}

\ifx\compilefullpaper\undefined  
\end{document}
\fi